\documentclass[ams,aps,prb,twocolumn,groupedaddress,showpacs,floatfix]{revtex4}
\usepackage{amssymb}

\usepackage{graphicx}
\usepackage{amsmath}

\begin{document}

\title{The Analytical Theory of Bulk Melting I: \\
Exact Solution of the One-dimensional Atom Chain}
\author{Yajun Zhou $\dagger$ and Xiaofeng Jin $\ddagger$}
\affiliation{Surface Physics Laboratory \& Department of Physics,
Fudan University, Shanghai 200433, China}
\date{\today}

\begin{abstract}
We investigate theoretically the crucial r\^{o}le of
interstitialcies that trigger the melting of a boundary-free
crystal. Based on an interstitialcy model that resembles the
$J_1$-$J_2$ model of frustrated antiferromagnets with uniaxial
anisotropy, we have calculated the exact partition function and
correlation functions in a one-dimensional atom chain. The melting
point and correlation behavior of this crystal model show the
applicability of Lindemann criterion and Born criterion in the
one-dimensional case.
\end{abstract}

\maketitle




\section{Introduction}

The physical picture of solid-liquid phase transition has emerged as a
controversial issue motivated by both science and industry since early 20th
century when divergent accounts of the instability mechanism and atom-scale
pathway towards melting in various theoretical models did not seem to yield
a unanimous prediction for the melting point of a realistic system \cite%
{Lindemann, Born, Att1, Att2, Att3, Att4, Lennard-Jones and Devonshire}. The
widely-cited Lindemann \cite{Lindemann} and Born \cite{Born} criteria, do
not form an exception to this discrepancy at a glance: in the former
criterion, Lindemann proposed that melting is triggered by the avalanche of
the root-mean-square atom displacement as soon as it exceeds a threshold
fraction ($\delta _{L}^{\ast }$) of the atom spacing, where $\delta
_{L}^{\ast }$ is called the critical Lindemann ratio, a semi-empirical
parameter initially conceived as a lattice type characteristic (but
experiments \cite{Lindemann rule} suggest otherwise); in the latter
criterion, Born argued that the vanishment of shear modulus (Ref. \cite{Born
rule} challenges this vanishment experimentally) is responsible for the
inability to resist lattice destruction at the melting point.

In this series of papers, we give a detailed presentation of how to join the
Lindemann and Born criteria together. We use a model analogous to the $J_{1}$%
-$J_{2}$ model for frustrated antiferromagnets to formulate that the
cooperative creation and the spatial correlation of interstitialcies are the
impetus that triggers and propagates instability in a crystal and that
eventually undermines long-range order in the surface-free solid (bulk
material). This idea is inspired by a recent molecular dynamics simulation
in which the r\^{o}le of interstitialcies is emphasized \cite{Jin}, the
previous understandings of the r\^{o}les of vacancies in surface melting %
\cite{surface melting}, the formation and motion of interstitial clusters in
metals \cite{Baskes, Halpern, Sandberg, Self-interstitial} and
semiconductors \cite{Si-interstitial, Ge-interstitial, As-interstitial},
early theories of interstitialcies based on computer simulations \cite%
{Stilinger and Weber, Juelich} and some thermodynamic arguments \cite%
{first-order transition, Granato}, and especially the notion of ``virtual
attraction'' between defects in Ref. \cite{Stilinger and Weber}.

This paper deals with the exact solution of the boundary-free
one-dimensional (1D) atom chain where two independent analytical
algorithms are applied to evaluate the partition function.
Sect.~II presents a brief account of the methodology of our
interstitialcy model; Sect.~III\ provides an exact solution to the
model based on the transfer matrix; Sect.~IV is the summary of the
results in Paper~I; Appendix A elaborates on a solution to our
model through Kac-Ward method, which serves as a valuable check to
the results in Sect.~III.

\section{Model Hamiltonian}

We begin our argument with the Hamiltonian of a 1D atom chain
composed of $N$ atoms and $2N$ sites (See FIG.~1 for the site
labelling):
\begin{equation}
\mathcal{H}_{\text{conf}}=\sum_{k=1}^{2N}\left(
J_{1}n_{k}n_{k+1}+J_{2}n_{k}n_{k+2}\right) ,  \label{1DH}
\end{equation}%
where the Boolean variable $n_{k}$ is the number of atoms occupying site $k$%
. $n_{k}=0$ or $1$. In this paper, we assume $J_{1}>0$ and $J_{2}<0$ unless
specified. In this Hamiltonian, the potential energy is set to be $J_{1}$
for each nearest-neighbor (NN) atom pair, $J_{2}$ for each
next-nearest-neighbor (NNN) atom pair and zero otherwise. In order to adapt
to the question under investigation, we need two restrictions
\begin{equation}
n_{k+2N}=n_{k},\sum_{k=1}^{2N}n_{k}=N
\end{equation}%
to reflect the periodical boundary condition and atom number
conservation respectively. By using the transformation \cite{Yang
and Lee, Pathria, Huang} $\sigma _{k}=2n_{k}-1$, one could map the
Hamiltonian above to the $J_{1}$-$J_{2}$ model \cite{Frustrated
magnets} of a 1D frustrated antiferromagnet
\begin{equation}
\mathcal{H}_{\text{conf}}=\frac{1}{4}\sum_{k=1}^{2N}\left( J_{1}\sigma
_{k}\sigma _{k+1}+J_{2}\sigma _{k}\sigma _{k+2}\right) +\mathrm{const},
\end{equation}%
where $\sigma _{k}=\sigma _{k+2N}=\pm 1$, and
$\sum_{k=1}^{2N}\sigma _{k}=0$. When $T=0$K, sites labeled by an
odd (or even) number $k$ are all occupied by atoms. In the light
of this, we may call the odd-number sites as lattice sites and
call the even-number sites as interstitialcy sites, as shown in
FIG.~1, or vice versa. The lattice sites and interstitialcy sites
interpenetrate.

\begin{figure}[t]
\center{\includegraphics[width=8cm]{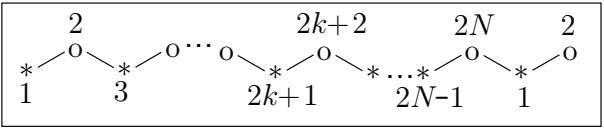}}
\caption{The way we label the lattice sites ``$\ast $'' and interstitialcy
sites ``o'' in 1D atom chain.}
\label{fig1:a}
\end{figure}

\section{The transfer matrix approach}

\subsection{The Partition Function for $T>0$K and Thermodynamic Properties}

To lift the constraint $\sum_{k=1}^{2N}\sigma _{k}=0$, we introduce an
phenomenological external field $h$ and consider the following Hamiltonian:
\begin{equation}
\mathcal{H}\left( \sigma ,h\right) =h\sum_{k=1}^{2N}\sigma _{k}+\frac{1}{4}%
\sum_{k=1}^{2N}\left( J_{1}\sigma _{k}\sigma _{k+1}+J_{2}\sigma _{k}\sigma
_{k+2}\right)
\end{equation}%
with the condition $\sigma _{k+2N}=\sigma _{k}$. If one could (i) evaluate
the canonical partition function corresponding to $\mathcal{H}\left( \sigma
,0\right) $ for any absolute temperature $T>0$K (to be elaborated in this
subsection):
\begin{eqnarray}
Q_{N} &=&\mathrm{Tr}e^{-\mathcal{H}\left( \sigma ,0\right)
/k_{B}T}  \notag
\\
&=&\sum_{\left\{ \sigma _{k}=\pm 1\right\} }e^{-\mathcal{H}\left( \sigma
,0\right) /k_{B}T}
\end{eqnarray}%
($k_{B}$: Boltzmann constant); and (ii) could show that the ``ensemble
average''%
\begin{eqnarray}
&&\left\langle \frac{1}{N}\sum_{k=1}^{2N}\sigma _{k}\right\rangle _{\mathcal{%
H}\left( \sigma ,0\right) }  \notag \\
&=&\left. \frac{1}{N}\frac{\partial }{\partial h}\mathrm{Tr}e^{-\mathcal{H}%
\left( \sigma ,h\right) /k_{B}T}\right| _{h=0}  \notag \\
&=&\mathrm{Tr}\left[ \left( \frac{1}{N}\sum_{k=1}^{2N}\sigma _{k}\right) e^{-%
\mathcal{H}\left( \sigma ,0\right) /k_{B}T}\right]   \label{TTrS}
\end{eqnarray}%
vanishes in the thermodynamic limit $N\rightarrow \infty $ (to be elaborated
in the next subsection), then it will be safe to say that $Q_{N}=\mathrm{Tr}%
e^{-\mathcal{H}\left( \sigma ,0\right) /k_{B}T}$ is also the exact partition
function corresponding to the Hamiltonian $\mathcal{H}_{\text{conf}}$ in the
thermodynamic limit because $\left\langle \frac{1}{N}\sum_{k=1}^{2N}\sigma
_{k}\right\rangle _{\mathcal{H}_{\text{conf}}}\equiv 0.$

In order to evaluate $Q_{N}=\mathrm{Tr}e^{-\mathcal{H}\left(
\sigma
,0\right) /k_{B}T}$, we need to construct a $4\times 4$ transfer matrix $%
\widehat{T}=\left( \widehat{T}_{\Sigma \Sigma ^{^{\prime }}}\right) $ so
that \cite{Nigel Goldenfeld, 2DIsing}
\begin{eqnarray}
Q_{N} &=&\sum_{\left( \sigma \right) }\exp \left[ -\frac{1}{4k_{B}T}%
\sum_{k=1}^{2N}\left( J_{1}\sigma _{k}\sigma _{k+1}+J_{2}\sigma _{k}\sigma
_{k+2}\right) \right]   \notag \\
&=&\sum_{\left( \sigma \right) }\widehat{T}_{\Sigma _{1}\Sigma _{2}}\widehat{%
T}_{\Sigma _{2}\Sigma _{3}}\cdots \widehat{T}_{\Sigma _{N-1}\Sigma _{N}}=%
\mathrm{Tr}\widehat{T}^{N}
\end{eqnarray}%
where $\Sigma _{k}=\sigma _{2k-1}+i\sigma _{2k}$. Such a transfer matrix
must take the form:

\begin{eqnarray}
&&\widehat{T}  \notag \\
&=&\left[
\begin{array}{llll}
\widehat{T}_{-1-i,-1-i} & \widehat{T}_{-1-i,-1+i} & \widehat{T}_{-1-i,1-i} &
\widehat{T}_{-1-i,1+i} \\
\widehat{T}_{-1+i,-1-i} & \widehat{T}_{-1+i,-1+i} & \widehat{T}_{-1+i,1-i} &
\widehat{T}_{-1+i,1+i} \\
\widehat{T}_{1-i,-1-i} & \widehat{T}_{1-i,-1+i} & \widehat{T}_{1-i,1-i} &
\widehat{T}_{1-i,1+i} \\
\widehat{T}_{1+i,-1-i} & \widehat{T}_{1+i,-1+i} & \widehat{T}_{1+i,1-i} &
\widehat{T}_{1+i,1+i}%
\end{array}%
\right]   \notag \\
&=&\left[
\begin{array}{llll}
e^{-2\beta ^{\prime }\left( J_{1}+J_{2}\right) } & e^{-\beta ^{\prime }J_{1}}
& e^{\beta ^{\prime }J_{1}} & e^{2\beta ^{\prime }J_{2}} \\
e^{\beta ^{\prime }J_{1}} & e^{2\beta ^{\prime }\left( J_{1}-J_{2}\right) }
& e^{2\beta ^{\prime }J_{2}} & e^{-\beta ^{\prime }J_{1}} \\
e^{-\beta ^{\prime }J_{1}} & e^{2\beta ^{\prime }J_{2}} & e^{2\beta ^{\prime
}\left( J_{1}-J_{2}\right) } & e^{\beta ^{\prime }J_{1}} \\
e^{2\beta ^{\prime }J_{2}} & e^{\beta ^{\prime }J_{1}} & e^{-\beta ^{\prime
}J_{1}} & e^{-2\beta ^{\prime }\left( J_{1}+J_{2}\right) }%
\end{array}%
\right]   \notag \\
&=&\left[
\begin{array}{llll}
\frac{\left( 1-x_{1}\right) \left( 1-x_{2}\right) }{\left( 1+x_{1}\right)
\left( 1+x_{2}\right) } & \sqrt{\frac{1-x_{1}}{1+x_{1}}} & \sqrt{\frac{%
1+x_{1}}{1-x_{1}}} & \frac{1+x_{2}}{1-x_{2}} \\
\sqrt{\frac{1+x_{1}}{1-x_{1}}} & \frac{\left( 1+x_{1}\right) \left(
1-x_{2}\right) }{\left( 1-x_{1}\right) \left( 1+x_{2}\right) } & \frac{%
1+x_{2}}{1-x_{2}} & \sqrt{\frac{1-x_{1}}{1+x_{1}}} \\
\sqrt{\frac{1-x_{1}}{1+x_{1}}} & \frac{1+x_{2}}{1-x_{2}} & \frac{\left(
1+x_{1}\right) \left( 1-x_{2}\right) }{\left( 1-x_{1}\right) \left(
1+x_{2}\right) } & \sqrt{\frac{1+x_{1}}{1-x_{1}}} \\
\frac{1+x_{2}}{1-x_{2}} & \sqrt{\frac{1+x_{1}}{1-x_{1}}} & \sqrt{\frac{%
1-x_{1}}{1+x_{1}}} & \frac{\left( 1-x_{1}\right) \left( 1-x_{2}\right) }{%
\left( 1+x_{1}\right) \left( 1+x_{2}\right) }%
\end{array}%
\right]   \notag \\
&&
\end{eqnarray}%
where $\beta ^{\prime }=\left( 4k_{B}T\right) ^{-1}$ and $x_{1}=\tanh \beta
^{\prime }J_{1}$, $x_{2}=\tanh \beta ^{\prime }J_{2}$. The matrix elements
are obtained like the following example:
\begin{widetext}
\begin{eqnarray}
& &\widehat{T}_{1-i,-1+i}=\exp \left\{ -\beta ^{\prime }\bpm
\includegraphics[height=0.73in]{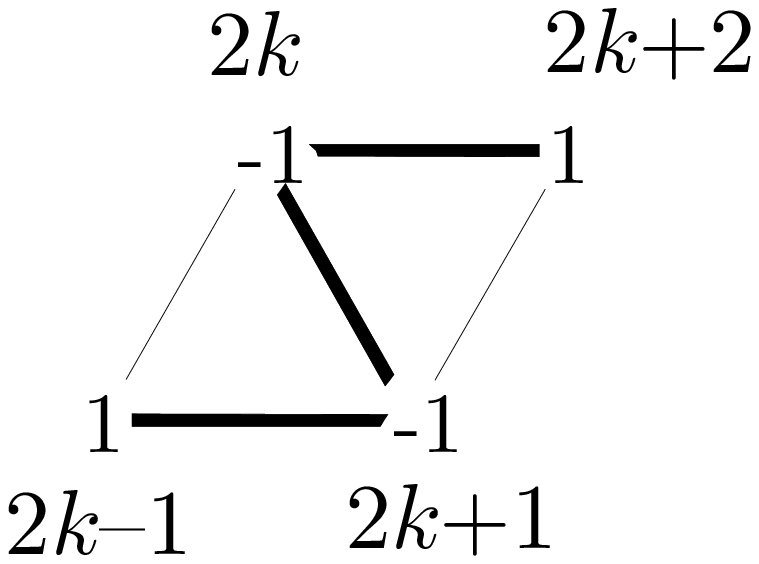}\epm \right\}\notag=\exp \left\{ -\beta ^{\prime }\bpm
\includegraphics[height=0.73in]{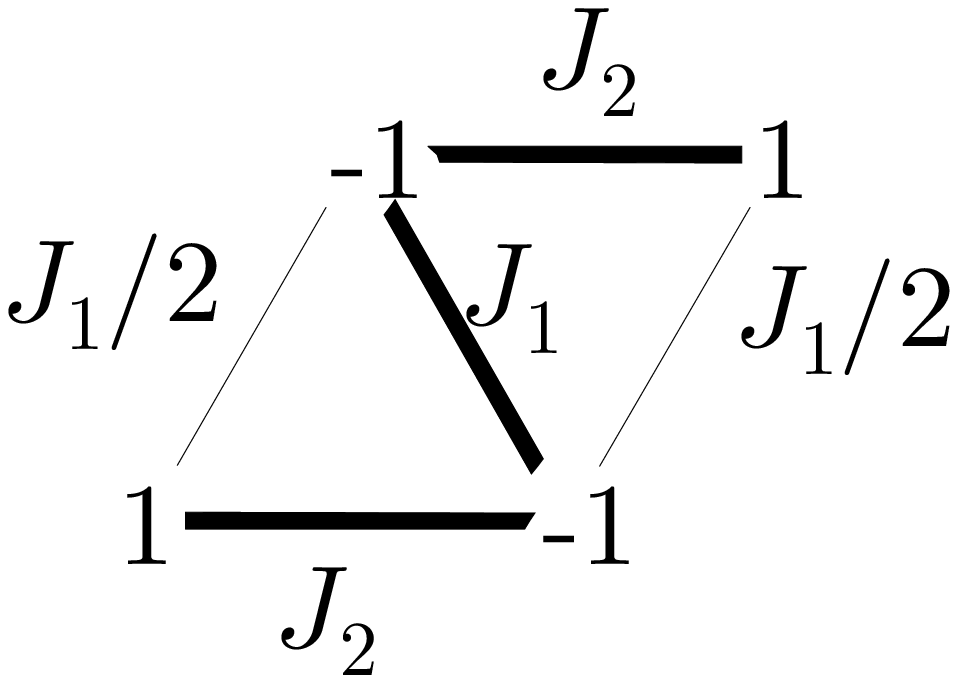}\epm \right\} \notag\\
&=&\exp \{ -\beta ^{\prime }[ \left( -1\right) \times \left(
1\right) \frac{J_1}2+(-1)\times \left( 1\right) J_2+\left(
-1\right) \times \left( -1\right) \frac{J_1}2+(-1)\times \left(
1\right) J_2+\left( -1\right)
\times \left( 1\right) \frac{J_1}2] \}\notag \\
&=&\exp \left( 2\beta ^{\prime }J_2\right)%
\end{eqnarray}
\end{widetext} Each interior interaction (thick black edges) energy in the
cell diagram is counted as one, while each peripheral interaction (thin
black edges) is counted as half because such ``bonds'' are shared by two
cells.

The transfer matrix is diagonalized by $\widehat{T}=\widehat{S}\widehat{L}%
\widehat{S}^{-1}$, where $\widehat{L}=\mathrm{diag}\left[ \lambda
_{1},\lambda _{2},\lambda _{3},\lambda _{4}\right] $ is a matrix with four
eigenvalues of $\widehat{T}$ as its diagonal elements. Therefore, $%
Q_{N}=\lambda _{1}^{N}+\lambda _{2}^{N}+\lambda _{3}^{N}+\lambda
_{4}^{N}$. Here,
\begin{widetext}
\begin{equation}
\begin{array}{c}
\lambda _{1} \\
\lambda _{2}%
\end{array}%
=\frac{2\left[ 1+x_{2}^{2}-2x_{2}x_{1}^{2}\mp \left( 1-x_{2}\right) \sqrt{%
\left( 1+x_{2}\right) ^{2}-4x_{2}x_{1}^{2}}\right] }{\left(
1-x_{1}^{2}\right) \left( 1-x_{2}^{2}\right) },
\end{equation}

\begin{equation}
\begin{array}{c}
\lambda _{3} \\
\lambda _{4}%
\end{array}%
=\frac{2\left[ x_{2}\left( x_{2}x_{1}^{2}-2\right) +x_{1}^{2}\mp
\left(
1-x_{2}\right) \sqrt{\left( 1+x_{2}\right) ^{2}x_{1}^{2}-4x_{2}}\right] }{%
\left( 1-x_{1}^{2}\right) \left( 1-x_{2}^{2}\right) }.
\end{equation}

\end{widetext} In the thermodynamic limit, only the largest eigenvalue $%
\lambda _{2}$ survives in the final expression of $\frac{1}{N}\log
Q_{N}\rightarrow \log \lambda _{2}$. Hence we have the free energy $F_{\text{%
conf}}$ as shown below:%
\begin{eqnarray}
&&F_{\text{conf}}\left( J_1,J_2,T\right)   \notag \\
&=&Nk_{B}T\{\log \left[ \left( 1-x_{1}^{2}\right) \left( 1-x_{2}^{2}\right) %
\right] -\log 2  \notag \\
&&-\log [1+x_{2}^{2}-2x_{2}x_{1}^{2}  \notag \\
&&+\allowbreak \left( 1-x_{2}\right) \sqrt{\left( 1+x_{2}\right)
^{2}-4x_{2}x_{1}^{2}}]\},  \label{F}
\end{eqnarray}%
in the thermodynamic limit.

From this configurational free energy $F_{\text{conf}}$ evaluated from the
partition function, which is an analytical function for all $J_{1}$, $J_{2}$
when $T>0\mathrm{K}$, one may draw the following inferences:

(1) The occurrence of phase transition at finite temperature is
ruled out
because the partition function of the system, which reads $e^{-F_{\text{conf}%
}/k_{B}T}$, encounters no singularities (as visualized in FIG.~2) in the thermodynamic limit when $T>0%
\mathrm{K}$ because the argument of the logarithm never vanishes
in such cases. In other words, the system has only one phase for
any non-zero absolute temperature.

(2) The configurational entropy $S_{\text{conf}}$ could be evaluated by%
\begin{eqnarray}
\frac{S_{\text{conf}} }{Nk_{B}} &=&-\frac{1}{Nk_{B}}\frac{%
\partial F_{\text{conf}}}{\partial T}  \notag \\
&=&-\frac{1}{4x_{2}}\{(1+x_{2}^{2})\frac{J_{2}}{k_{B}T}+\left(
1-x_{2}\right) \times  \notag \\
&&[-\left( 1+x_{2}\right) ^{2}\frac{J_{2}}{k_{B}T}+2x_{2}x_{1}\frac{J_{1}}{%
k_{B}T}]\times  \notag \\
&&[\left( 1+x_{2}\right) ^{2}-4x_{2}x_{1}^{2}]^{-1/2}\}+\log 2  \notag \\
&&-\log \left[ \left( 1-x_{1}^{2}\right) \left( 1-x_{2}^{2}\right) \right]
\notag \\
&&+\log [1+x_{2}^{2}-2x_{2}x_{1}^{2}  \notag \\
&&+\allowbreak \left( 1-x_{2}\right) \sqrt{\left( 1+x_{2}\right)
^{2}-4x_{2}x_{1}^{2}}].
\end{eqnarray}

(3) The short-range correlation functions $\left\langle \sigma _{k}\sigma
_{k+1}\right\rangle $ and $\left\langle \sigma _{k}\sigma
_{k+2}\right\rangle $ ($\left\langle \cdot \right\rangle $: ensemble
average) read:

\begin{eqnarray}
\left\langle \sigma _{k}\sigma _{k+1}\right\rangle &=&\frac{4k_BT}{N}\frac{%
\partial }{\partial J_{1}}\left( \frac{1}{2}F_{\text{conf}}\right)  \notag \\
&=&-\frac{\left( 1-x_{2}\right) x_{1}}{\sqrt{\left( 1+x_{2}\right)
^{2}-4x_{2}x_{1}^{2}}} \\
\left\langle \sigma _{k}\sigma _{k+2}\right\rangle &=&\frac{4k_BT}{N}\frac{%
\partial }{\partial J_{2}}\left( \frac{1}{2}F_{\text{conf}}\right)  \notag \\
&=&-\frac{1+x_{2}^{2}-\frac{\left( 1-x_{2}^{2}\right) \left( 1+x_{2}\right)
}{\sqrt{\left( 1+x_{2}\right) ^{2}-4x_{2}x_{1}^{2}}}}{2x_{2}}
\end{eqnarray}

(4) $S_{\text{conf,}}\left\langle \sigma _{k}\sigma _{k+1}\right\rangle $
and $\left\langle \sigma _{k}\sigma _{k+2}\right\rangle $ changes
continuously as the temperature approaches absolute zero.

The equation%
\begin{equation}
S_{\text{conf}}\left(J_1,J_2,T\right) \rightarrow 0\text{, as }T\rightarrow 0%
\mathrm{K}\text{\textrm{,}}
\end{equation}%
is in accordance with the third law of thermodynamics; the equations \cite%
{J1}%
\begin{eqnarray}
\left\langle \sigma _{k}\sigma _{k+1}\right\rangle  &\rightarrow &-1\text{,
as }T\rightarrow 0\mathrm{K;} \\
\left\langle \sigma _{k}\sigma _{k+2}\right\rangle  &\rightarrow &+1\text{,
as }T\rightarrow 0\mathrm{K.}
\end{eqnarray}%
agree with the intuitive picture at absolute zero: NN sites should contain
one atom fixed at one site and one vacancy at the other (hence $\left\langle
n_{k}n_{k+1}\right\rangle =0$, $\left\langle \sigma _{k}\sigma
_{k+1}\right\rangle =-1$ for all $k$); NNN sites should be both occupied or
both vacant (hence $\left\langle n_{2k}n_{2k+2}\right\rangle =0$, $%
\left\langle n_{2k-1}n_{2k+1}\right\rangle =$ $1$, $\left\langle \sigma
_{k}\sigma _{k+2}\right\rangle =+1$ for all $k$).

\begin{figure}[b]
\center{\includegraphics[width=8cm]{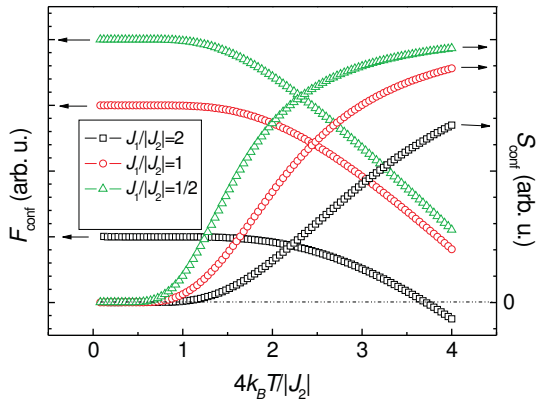} \caption{\label{fig2:b}
This plots configurational free energy $F_{\text{conf}}$ and
entropy $S_{\text{conf}}$ versus temperature for various $J_1/J_2$
ratios. $S_{\text{conf}}$=0 at 0K in all these cases.}}
\end{figure}

\subsection{Ensemble Expectation and Correlations via the Transfer Matrix}

Now we will employ the transfer matrix \cite{Nigel Goldenfeld,
2DIsing} \ to prove that the ensemble expectation $\left\langle
\Sigma _{k}\right\rangle $ vanishes in the thermodynamic limit for
all $k$ at finite temperature, which infers the vanishment of the
right side of Eqn.~(\ref{TTrS}). This will not only justify our
procedure of lifting the ``atom-number conservation'' constraint
but also show that the long-range order is non-existent at finite
temperature.

\begin{figure}[b]
\center{\includegraphics[width=8cm]{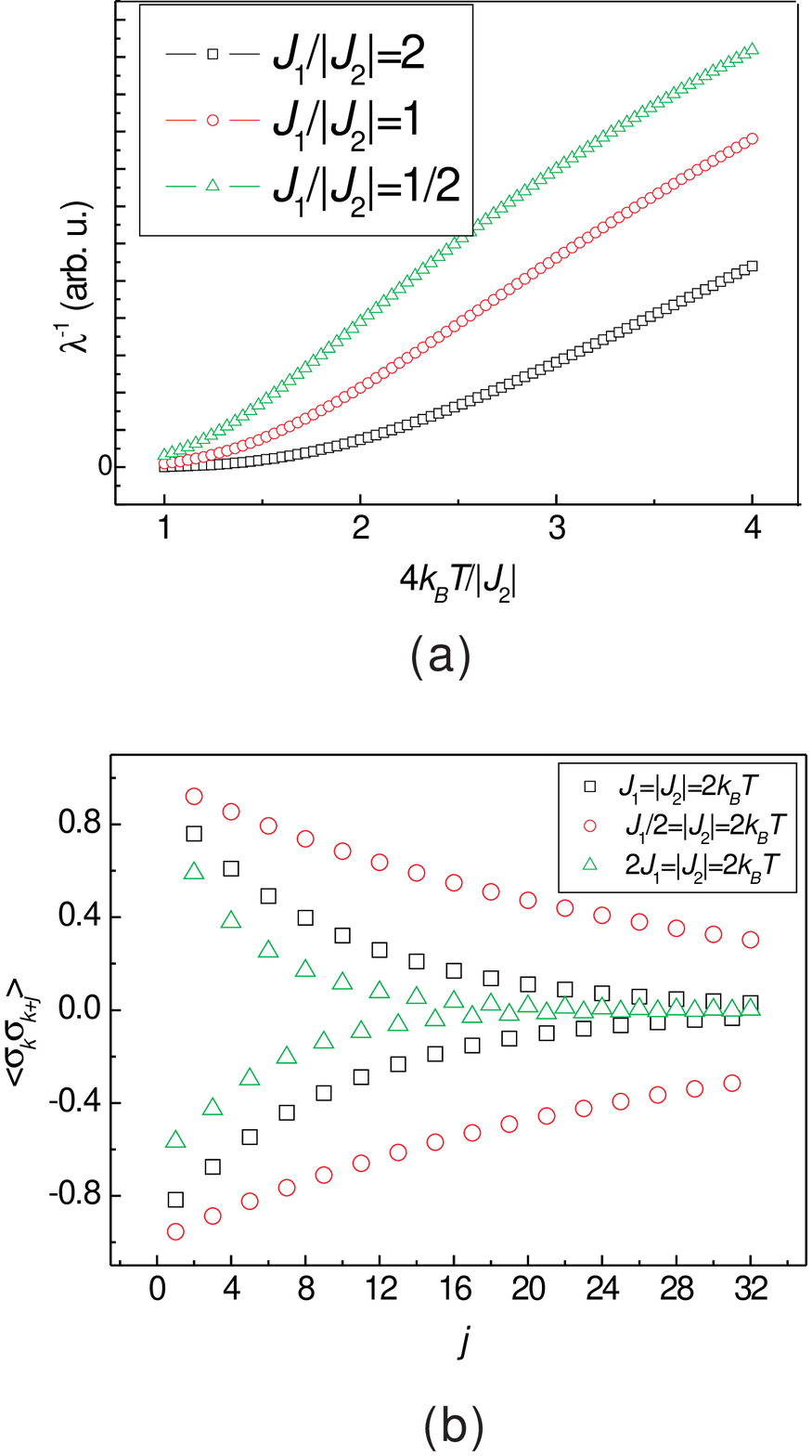}}
\caption{(a) The inverse correlation length $\protect\lambda ^{-1}$ versus
temperature for various $J_1/J_2$ ratios. Note that as $T$ approaches
absolute zero, so does $\protect\lambda ^{-1}$ because the correlation
length is $\infty$ at 0K. (b) The correlation function $\left\langle \protect%
\sigma _{k}\protect\sigma _{k+j}\right\rangle $ versus $j$ for fixed
temperature and $J_1/J_2$ ratios.}
\label{fig3:c}
\end{figure}

We find that

\begin{eqnarray}
\left\langle \Sigma _{k}\right\rangle
&=&\frac{1}{Q_{N}}\sum_{\left( \sigma
\right) }\Sigma _{k}\times  \notag \\
&&\exp \left[ -\beta ^{\prime }\sum_{k=1}^{2N}\left( J_{1}\sigma _{k}\sigma
_{k+1}+J_{2}\sigma _{k}\sigma _{k+2}\right) \right]  \notag \\
&=&\frac{1}{Q_{N}}\sum_{\left( \sigma \right) }\widehat{T}_{\Sigma
_{1}\Sigma _{2}}\widehat{T}_{\Sigma _{2}\Sigma _{3}}\cdots  \notag \\
&&\cdots \left( \widehat{T}_{\Sigma _{k-1}\Sigma _{k}}\Sigma _{k}\widehat{T}%
_{\Sigma _{k}\Sigma _{k+1}}\right) \cdots \widehat{T}_{\Sigma _{N-1}\Sigma
_{N}}  \notag \\
&=&\frac{1}{Q_{N}}\sum_{\left( \sigma \right) }\widehat{T}_{\Sigma
_{1}\Sigma _{2}}\widehat{T}_{\Sigma _{2}\Sigma _{3}}\cdots  \notag \\
&&\cdots \left( \widehat{A}_{\Sigma _{k-1}\Sigma _{k+1}}\right) \cdots
\widehat{T}_{\Sigma _{N-1}\Sigma _{N}}.
\end{eqnarray}

Here, $\widehat{A}=\widehat{T}\left( \widehat{\sigma }_{z}\otimes \widehat{1}%
+i\widehat{1}\otimes \widehat{\sigma }_{z}\right) \widehat{T}$, $\widehat{%
\sigma }_{z}$ is the Pauli matrix and
\begin{eqnarray}
&&\widehat{\sigma }_{z}\otimes \widehat{1}+i\widehat{1}\otimes \widehat{%
\sigma }_{z}  \notag \\
&=&\left[
\begin{array}{cccc}
1+i & 0 & 0 & 0 \\
0 & 1-i & 0 & 0 \\
0 & 0 & -1+i & 0 \\
0 & 0 & 0 & -1-i%
\end{array}%
\right] .
\end{eqnarray}%
Therefore,
\begin{eqnarray}
\left\langle \Sigma _{k}\right\rangle
&=&\frac{1}{Q_{N}}\mathrm{Tr}\left[
\left( \widehat{\sigma }_{z}\otimes \widehat{1}+i\widehat{1}\otimes \widehat{%
\sigma }_{z}\right) \widehat{T}^{N}\right]  \notag \\
&=&\frac{1}{Q_{N}}\mathrm{Tr}\left[ \widehat{S}^{-1}\left( \widehat{\sigma }%
_{z}\otimes \widehat{1}+i\widehat{1}\otimes \widehat{\sigma }_{z}\right)
\widehat{S}\widehat{L}^{N}\right] .  \notag \\
&&
\end{eqnarray}%
As a matter of fact, $\widehat{S}^{-1}( \widehat{\sigma
}_{z}\otimes \widehat{1}+i\widehat{1}\otimes \widehat{\sigma
}_{z}) \widehat{S}$ takes the shape of
\begin{equation}
\left[
\begin{array}{llll}
0 & 0 & a & b \\
0 & 0 & c & d \\
e & f & 0 & 0 \\
g & h & 0 & 0%
\end{array}%
\right]
\end{equation}%
where the eight Latin letters represent the elements that are not
necessarily zero.
\begin{eqnarray}
&&\mathrm{Tr}\left[ \widehat{S}^{-1}\left( \widehat{\sigma }_{z}\otimes
\widehat{1}+i\widehat{1}\otimes \widehat{\sigma }_{z}\right) \widehat{S}%
\widehat{L}^{N}\right]  \notag \\
&=&\mathrm{Tr}\left[
\begin{array}{cccc}
0 & 0 & a\lambda _{3}^{N} & b\lambda _{4}^{N} \\
0 & 0 & c\lambda _{3}^{N} & d\lambda _{4}^{N} \\
e\lambda _{1}^{N} & f\lambda _{2}^{N} & 0 & 0 \\
g\lambda _{1}^{N} & h\lambda _{2}^{N} & 0 & 0%
\end{array}%
\right] =0.
\end{eqnarray}%
It is now evident that in the limit that $N\rightarrow \infty $, $%
\left\langle \Sigma _{k}\right\rangle \rightarrow 0$, which is equivalent to
\begin{equation}
\sum_{j=1}^{N}\sigma _{2j}=\sum_{j=1}^{N}\sigma _{2j-1}=0,
\label{equal}
\end{equation}%
which is an indication of vanishing long-range order at finite
temperature because Eqn.~(\ref{equal}) suggests that the
interstitialcy sites are occupied by $N/2$ atoms and so are the
lattice sites.

From this result, we may conclude that unlike the short-range parameters
such as $\left\langle \sigma _{k}\sigma _{k+1}\right\rangle $ and $%
\left\langle \sigma _{k}\sigma _{k+2}\right\rangle $, the long-range order
parameter $\mathfrak{L}$ witnesses a catastrophe at 0K:

\begin{eqnarray}
L &\equiv& \frac{1}{2N}\sum_{k=1}^{2N}\left( -1\right)
^{k}\left\langle 2n_{k}-1\right\rangle=\left\{
\begin{array}{lll}
\pm 1, &  & T=0\mathrm{K} \\
0, &  & T>0\mathrm{K}%
\end{array}%
\right.
\end{eqnarray}%
Therefore, there exists a solid-liquid phase transition at
$0\mathrm{K}$, a process characterized by destruction of
long-range order and preservation of short-range order in the
meantime.

To obtain the correlation between $\Sigma _{k}$ at different
sites, we may extend the matrix method to the correlation function
$\Gamma \left( k,k+j\right) $ (for simplicity, $j$ is assumed to
be positive hereinafter):

\begin{widetext}
\begin{eqnarray}
&&\Gamma \left( k,k+j\right)  \notag \\
&\equiv&\left\langle \left( \Sigma _{k}-\left\langle \Sigma
_{k}\right\rangle \right) \left( \Sigma _{k+j}-\left\langle \Sigma
_{k+j}\right\rangle \right)
\right\rangle =\left\langle \Sigma _{k}\Sigma _{k+j}\right\rangle  \notag \\
&=&\frac{1}{Q_{N}}\mathrm{Tr}\left[ \widehat{S}^{-1}\left( \widehat{\sigma }%
_{z}\otimes \widehat{1}+i\widehat{1}\otimes \widehat{\sigma
}_{z}\right)
\widehat{S}\widehat{L}^{j}\widehat{S}^{-1}\left( \widehat{\sigma }%
_{z}\otimes \widehat{1}+i\widehat{1}\otimes \widehat{\sigma
}_{z}\right)
\widehat{S}\widehat{L}^{N-j}\right]  \notag \\
&=&\frac{1}{Q_{N}}\times  \notag \\
&&\mathrm{Tr}\left\{ \left[
\begin{array}{cc}
\left[ ae\left( \frac{\lambda _{3}}{\lambda _{1}}\right) ^{j}+bg\left( \frac{%
\lambda _{4}}{\lambda _{1}}\right) ^{j}\right] \lambda _{1}^{N} &
\left[
af\left( \frac{\lambda _{3}}{\lambda _{2}}\right) ^{j}+bh\left( \frac{%
\lambda _{4}}{\lambda 2}\right) ^{j}\right] \lambda _{2}^{N} \\
\left[ ce\left( \frac{\lambda _{3}}{\lambda _{1}}\right) ^{j}+dg\left( \frac{%
\lambda _{4}}{\lambda _{1}}\right) ^{j}\right] \lambda _{1}^{N} &
\left[
cf\left( \frac{\lambda _{3}}{\lambda _{2}}\right) ^{j}+dh\left( \frac{%
\lambda _{4}}{\lambda 2}\right) ^{j}\right] \lambda _{2}^{N}%
\end{array}%
\right] \right.  \notag \\
&&\left. \oplus \left[
\begin{array}{cc}
\left[ ae\left( \frac{\lambda _{1}}{\lambda _{3}}\right) ^{j}+cf\left( \frac{%
\lambda _{2}}{\lambda _{3}}\right) ^{j}\right] \lambda _{3}^{N} &
\left[
be\left( \frac{\lambda _{1}}{\lambda _{4}}\right) ^{j}+df\left( \frac{%
\lambda _{2}}{\lambda _{4}}\right) ^{j}\right] \lambda _{4}^{N} \\
\left[ ag\left( \frac{\lambda _{1}}{\lambda _{3}}\right) ^{j}+ch\left( \frac{%
\lambda _{2}}{\lambda _{3}}\right) ^{j}\right] \lambda _{3}^{N} &
\left[
bg\left( \frac{\lambda _{1}}{\lambda _{4}}\right) ^{j}+dh\left( \frac{%
\lambda _{2}}{\lambda _{4}}\right) ^{j}\right] \lambda _{4}^{N}%
\end{array}%
\right] \right\}  \notag \\
&\rightarrow &cf\left( \frac{\lambda _{3}}{\lambda _{2}}\right)
^{j}+dh\left( \frac{\lambda _{4}}{\lambda _{2}}\right) ^{j}=cfe^{-\frac{2j}{%
\mu }}+dhe^{-\frac{2j}{\lambda }}.
\end{eqnarray}%
where

\begin{equation}
\frac{cf}i=\frac{\left( 1-x_2\right) \left[ \sqrt{\left(
1+x_2\right)
^2-4x_2x_1^2}-x_1\sqrt{\left( 1+x_2\right) ^2x_1^2-4x_2}\right] }{\sqrt{%
\left( 1+x_2\right) ^2x_1^2-4x_2}\sqrt{\left( 1+x_2\right)
^2-4x_2x_1^2}},
\end{equation}
\newline
\begin{equation}
\frac{dh}i=-\frac{\left( 1-x_2\right) \left[ \sqrt{\left(
1+x_2\right)
^2-4x_2x_1^2}+x_1\sqrt{\left( 1+x_2\right) ^2x_1^2-4x_2}\right] }{\sqrt{%
\left( 1+x_2\right) ^2x_1^2-4x_2}\sqrt{\left( 1+x_2\right)
^2-4x_2x_1^2}},
\end{equation}
and $\mu \ll \lambda$,

\begin{eqnarray}
\mu &=& 2\left( \log \frac{\lambda _2}{\lambda
_3}\right)^{-1}=2\left[ \log \frac{1+x_2^2-2x_2x_1^2+\left(
1-x_2\right) \sqrt{\left( 1+x_2\right) ^2-4x_2x_1^2}}{x_1^2-\left(
1-x_2\right) x_1\sqrt{\left(
1+x_2\right) ^2x_1^2-4x_2}-x_2\left( 2-x_2x_1^2\right) }\right] ^{-1}, \\
\lambda &=&2\left( \log \frac{\lambda _2}{\lambda _4}\right)^{-1}=2\left[ \log \frac{1+x_2^2-2x_2x_1^2+\left( 1-x_2\right) \sqrt{%
\left( 1+x_2\right) ^2-4x_2x_1^2}}{x_1^2+\left( 1-x_2\right)
x_1\sqrt{\left( 1+x_2\right) ^2x_1^2-4x_2}-x_2\left(
2-x_2x_1^2\right) }\right] ^{-1}.
\end{eqnarray}

\end{widetext}With the definition of $\Sigma _{k}$, we obtain that
\begin{eqnarray}
\left\langle \Sigma _{k}\Sigma _{k+j}\right\rangle  &=&\left( \left\langle
\sigma _{2k-1}\sigma _{2k+2j-1}\right\rangle -\left\langle \sigma
_{2k}\sigma _{2k+2j}\right\rangle \right)   \notag \\
&&+i\left( \left\langle \sigma _{2k-1}\sigma _{2k+2j}\right\rangle
+\left\langle \sigma _{2k}\sigma _{2k+2j-1}\right\rangle \right)   \notag \\
&&
\end{eqnarray}%
Therefore, $\left\langle \sigma _{2k-1}\sigma _{2k+2j-1}\right\rangle
-\left\langle \sigma _{2k}\sigma _{2k+2j}\right\rangle =0$ (translational
invariance) and
\begin{eqnarray}
&&G\left( 0,2j+1\right) +G\left( 0,2j-1\right)   \notag \\
&=&\left\langle \sigma _{2k-1}\sigma _{2k+2j}\right\rangle +\left\langle
\sigma _{2k}\sigma _{2k+2j-1}\right\rangle   \notag \\
&=&\frac{cf}{i}\left( \frac{\lambda _{3}}{\lambda _{2}}\right) ^{j}+\frac{dh%
}{i}\left( \frac{\lambda _{4}}{\lambda _{2}}\right) ^{j}.
\end{eqnarray}%
Similarly, we explore the properties of $\Sigma _{k}^{\prime }=\sigma
_{2k-1}-\sigma _{2k}$ (obtained by replacing the factor $i$ in $\Sigma _{k}$
with $-1$) by the following procedures:

(1) $\widehat{S}^{-1}( \widehat{\sigma }_{z}\otimes \widehat{1}-%
\widehat{1}\otimes \widehat{\sigma }_{z}) \widehat{S}$ takes the
shape of
\begin{equation}
\left[
\begin{array}{llll}
0 & 0 & \alpha & \beta \\
0 & 0 & \chi & \delta \\
\varepsilon & \phi & 0 & 0 \\
\gamma & \eta & 0 & 0%
\end{array}%
\right] ;
\end{equation}%
where each Greek letter represents a non-vanishing element.

(2) Verify that

\begin{widetext}
\begin{eqnarray}
&&\left\langle \Sigma _{k}^{\prime }\Sigma _{k+j}^{\prime
}\right\rangle
\notag \\
&=&\left( \left\langle \sigma _{2k-1}\sigma
_{2k+2j-1}\right\rangle +\left\langle \sigma _{2k}\sigma
_{2k+2j}\right\rangle \right) -\left( \left\langle \sigma
_{2k-1}\sigma _{2k+2j}\right\rangle +\left\langle \sigma
_{2k}\sigma _{2k+2j-1}\right\rangle \right)  \notag \\
&=&2G\left( 0,2j\right) -\left[ G\left( 0,2j+1\right) +G\left( 0,2j-1\right) %
\right]  \notag \\
&=&\frac{1}{Q_{N}}\mathrm{Tr}\left[ \widehat{S}^{-1}\left(
\widehat{\sigma _{z}}\otimes \widehat{1}-\widehat{1}\otimes
\widehat{\sigma _{z}}\right)
\widehat{S}\widehat{L}^{j}\widehat{S}^{-1}\left( \widehat{\sigma _{z}}%
\otimes \widehat{1}-\widehat{1}\otimes \widehat{\sigma _{z}}\right) \widehat{%
S}\widehat{L}^{N-j}\right]  \notag \\
&\rightarrow &\chi \phi \left( \frac{\lambda _{3}}{\lambda
_{2}}\right) ^{j}+\delta \eta \left( \frac{\lambda _{4}}{\lambda
_{2}}\right) ^{j}.
\end{eqnarray}%
Then we combine the results above to find the recursion relations:

\begin{equation}
G\left( 0,2j\right) =\frac{1}{2}\left[ \left( \frac{cf}{i}+\chi
\phi \right)
\left( \frac{\lambda _{3}}{\lambda _{2}}\right) ^{j}+\left( \frac{dh}{i}%
+\delta \eta \right) \left( \frac{\lambda _{4}}{\lambda _{2}}\right) ^{j}%
\right];
\end{equation}%
\newline
\begin{equation}
G\left( 0,2j+1\right) +G\left( 0,2j-1\right) =\frac{cf}{i}\left( \frac{%
\lambda _{3}}{\lambda _{2}}\right) ^{j}+\frac{dh}{i}\left( \frac{\lambda _{4}%
}{\lambda _{2}}\right)^{j} ,
\end{equation}%
\newline
and the initial condition
\begin{equation}
G\left( 0,1\right) =\left\langle \sigma _{k}\sigma _{k+1}\right\rangle =-%
\frac{\left( 1-x_{2}\right) x_{1}}{\sqrt{\left( 1+x_{2}\right)
^{2}-4x_{2}x_{1}^{2}}}
\end{equation}%
which finally results in
\begin{eqnarray}
&&G\left( 0,j\right)=\left\langle \sigma _{k}\sigma _{k+j}\right\rangle  \notag \\
&=&\frac{1}{2}\left( 1-\left( -1\right) ^{j}\right) \left[
\frac{cf}{i\left(
\frac{\lambda _{3}}{\lambda _{2}}+1\right) }\left( \frac{\lambda _{3}}{%
\lambda _{2}}\right) ^{\frac{j+1}{2}}+\frac{dh}{i\left( \frac{\lambda _{4}}{%
\lambda _{2}}+1\right) }\left( \frac{\lambda _{4}}{\lambda _{2}}\right) ^{%
\frac{j+1}{2}}\right]  \notag \\
&&+\frac{1}{4}\left( 1+\left( -1\right) ^{j}\right) \left[ \left( \frac{cf}{i%
}+\chi \phi \right) \left( \frac{\lambda _{3}}{\lambda _{2}}\right) ^{\frac{j%
}{2}}+\left( \frac{dh}{i}+\delta \eta \right) \left( \frac{\lambda _{4}}{%
\lambda _{2}}\right) ^{\frac{j}{2}}\right] \notag \\
&=&\frac 12\left( 1-\left(
-1\right) ^j\right) \left( Ae^{-\frac{\left| j\right| +1}\mu }+Be^{-\frac{%
\left| j\right| +1}\lambda }\right) +\frac 14\left( 1-\left(
-1\right)
^j\right) \left( A^{\prime }e^{-\frac{\left| j\right| }\mu }+B^{\prime }e^{-%
\frac{\left| j\right| }\lambda }\right).
\end{eqnarray}

where

\begin{eqnarray}
A &=&\frac{cf}{i\left( \frac{\lambda _3}{\lambda _2}+1\right) }=\frac{\left( 1+x_2\right) ^2\left( 1-x_1^2\right) -\left( 1-x_2\right) %
\left[ x_1\sqrt{\left( 1+x_2\right) ^2x_1^2-4x_2}-\sqrt{\left(
1+x_2\right) ^2-4x_2x_1^2}\right] }{2\sqrt{\left( 1+x_2\right)
^2x_1^2-4x_2}\sqrt{\left(
1+x_2\right) ^2-4x_2x_1^2}}; \\
B &=&\frac{dh}{%
i\left( \frac{\lambda _4}{\lambda _2}+1\right) }=\frac{-\left( 1+x_2\right) ^2\left( 1-x_1^2\right) -\left( 1-x_2\right) %
\left[ x_1\sqrt{\left( 1+x_2\right) ^2x_1^2-4x_2}+\sqrt{\left(
1+x_2\right) ^2-4x_2x_1^2}\right] }{2\sqrt{\left( 1+x_2\right)
^2x_1^2-4x_2}\sqrt{\left(
1+x_2\right) ^2-4x_2x_1^2}} ;\\
A^{\prime } &=&\frac{cf}{i}+\chi \phi =1-\frac{\left( 1-x_2\right) ^2x_1}{\sqrt{\left(
1+x_2\right)
^2x_1^2-4x_2}\sqrt{\left( 1+x_2\right) ^2-4x_2x_1^2}} ;\\
B^{\prime } &=&\frac{dh}{i}+\delta \eta=1+\frac{\left(
1-x_2\right) ^2x_1}{\sqrt{\left( 1+x_2\right)
^2x_1^2-4x_2}\sqrt{\left( 1+x_2\right) ^2-4x_2x_1^2}}.
\end{eqnarray}
\end{widetext}

For large $\left| j\right| $, the $e^{-\left| j\right| /\lambda }$
terms dominate the correlation function $\left\langle \sigma
_{k}\sigma _{k+j}\right\rangle $, so $\lambda $ serves as the
correlation length that characterizes the maximum range of
information transfer in the 1D atom chain. (This is visualized in
FIG.~3)

\subsection{Discussions}

From the exact solutions of $F_{\text{conf}}$ and $\left\langle
\Sigma _{k}\right\rangle $, we find that the 1D atom chain
``melts'' at 0K. The catastrophe of long-range order and
continuous change of short-range order compare reasonably with a
real solid-liquid phase transition in three dimensions. This
melting point in 1D atom chain is in perfect agreement with the
Lindemann criterion in the 1D case, recalling the well-established
result that the mean displacement of a 1D elastic atom chain is
$\infty $ for any positive temperature \cite{Peierls}. Although in
the model Hamiltonian Eqn.~(\ref{1DH}), we do not explicitly take
vibrational energy into account, the melting point in our model
still coincides with the prediction based on vibrational
instability for two reasons: (1) the pair
interaction mode in our model (a repulsion $J_{1}$ for NN, an attraction $%
J_{2}$ for NNN, say) still forms a reasonable caricature of the
potential well in a solid. (2) the ``hopping'' between lattice
sites and interstitialcy sites, which is the mode of atom movement
in our model, mimics the quantized motion of atoms at low
temperature.

One may also check Born's scenario in our 1D exact solution. In principle,
there is no such a counterpart of the three-dimensional shear moduli in the
1D case. Nevertheless, for fixed $J_{2}$ (bonding energy), greater $J_{1}$
infers greater energy gap between the NN contact and NNN contact modes. In
other words, when the bonding energy $J_{2}$ and temperature $T$ is fixed,
increasing $J_{1}$ will add to the difficulty of creating interstitialcies,
thereby increasing the ``rigidity'' of the 1D atom chain. In the light of
this, we may define a dimensionless ``rigidity parameter'' as $\left(
J_{1}+\left| J_{2}\right| \right) /\left| J_{2}\right| $. It can be verified
analytically that
\begin{widetext}
\begin{eqnarray}
&&\left( \frac{\partial S_{\text{conf}}\left( J_{1},J_{2},T\right) }{%
\partial J_{1}}\right) _{J_{2},T}=\frac{\left( 1-x_{1}^{2}\right) }{4k_{B}T}\left( \frac{\partial S_{\text{%
conf}}\left( x_{1},x_{2}\right) }{\partial x_{1}}\right) _{x_{2}}  \notag \\
&=&\frac{\left( 1-x_{1}^{2}\right) }{4k_{B}T}\frac{2\left(
x_{2}-1\right) \left( x_{2}+1\right) ^{2}}{\sqrt{\left[ \left(
1+x_{2}\right) ^{2}-4x_{1}^{2}x_{2}\right] ^{3}}}\left( \tanh
^{-1}x_{1}-2x_{1}\tanh
^{-1}x_{2}\right)  \notag \\
&<&0,(0<-x_{2},x_{1}<1)
\end{eqnarray}%
and%
\begin{eqnarray}
&&\left( \frac{\partial \lambda \left( J_{1},J_{2},T\right) }{\partial J_{1}}%
\right) _{J_{2},T}=\frac{\left( 1-x_{1}^{2}\right) }{4k_{B}T}\left( \frac{\partial
\lambda
\left( x_{1},x_{2}\right) }{\partial J_{1}}\right) _{x_{2}}  \notag \\
&=&\lambda ^{2}\frac{2\left( 1-x_{2}\right) }{4k_{B}T}\left[ \frac{1}{\sqrt{%
x_{1}^{2}\left( 1+x_{2}\right) ^{2}-4x_{2}}}-\frac{1}{\sqrt{\frac{1}{%
x_{1}^{2}}\left( 1+x_{2}\right) ^{2}-4x_{2}}}\right]  \notag \\
&>&0.(0<-x_{2},x_{1}<1)
\end{eqnarray}%
\end{widetext}Physically speaking, these two inequalities are reasonable
because the increase of rigidity will help to restore some short
range order to the system, thereby lowering the entropy and
increasing the correlation length. (See FIG.~2 and FIG.~3(a) for
specific examples.)

The exact solution above provides insight for the 3D case in that
correlation between atom occupancies stands to be the common means to
propagate instability in systems, no matter what dimensionality. In a
proceeding paper, we will present a detailed analysis of how such
correlation undermines the long-range order in a fcc crystal. Some numerical
corollaries of this 1D model may also shed light on the melting point
formula to be established in the 3D case. \cite{3DJ1} One heuristic approach
to ``guess'' the melting point formula in 3D is to find the ``critical''
temperature $T_{c}$ in the 1D model where the heat capacity has a maximum,
and speculate that such a $T_{c}$ will correspond to a heat capacity
discontinuity in the 3D case. A lengthy expression for the heat capacity%
\begin{equation}
c=T\frac{\partial S_{\text{conf}}}{\partial T}
\end{equation}%
in our 1D model could be obtained analytically. FIG.~4 presents a
``phase diagram'' for the sign of $\partial c/\partial T$ based on
the analytical expression, in which the phase boundary (where
$\partial c/\partial T=0$) could be approximated by a simple
formula \cite{1DJ1}:%
\begin{equation}
\tanh \frac{\left| J_{1}\right| }{4k_{B}T_{c}}+\tanh \frac{\left|
J_{2}\right| }{4k_{B}T_{c}}\approx 0.88.
\end{equation}

\begin{figure}[t]
\center{\includegraphics[width=8cm]{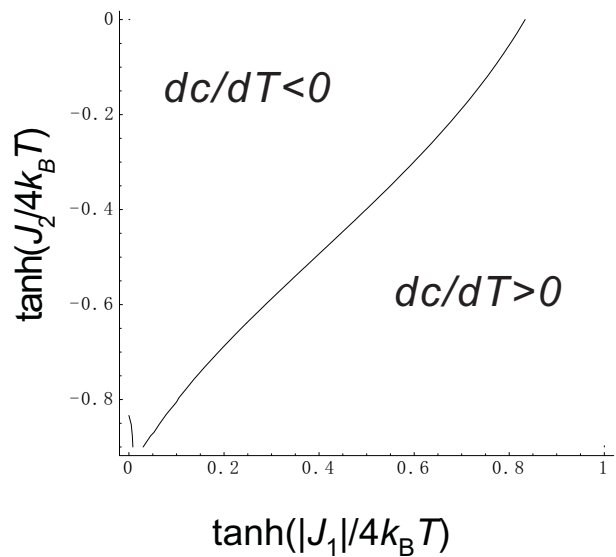}}
\caption{A ``phase diagram'' for $\partial c/\partial T$ ($c$: heat
capacity; $T$: absolute temperature) in which the phase boundary is
approximately a straight line.}
\label{fig4:d}
\end{figure}

When $\left| J_{1}\right| \approx \left| J_{2}\right| $, the formula above
could be replaced by
\begin{equation}
\left| J_{1}\right| +\left| J_{2}\right| \approx 3.52k_{B}T_{c}
\end{equation}%
without loss of precision. This suggests that in the scenario of 3D melting
at a temperature corresponding to such a $T_{c}$, the thermal energy $\sim
3k_{B}T_{c}$ is roughly tantamount to the energy barrier of interstitialcy
creation, or in other words, the ``rigidity'' is combatted by thermal
motion. As long as such a positive $T_{c}$ is existent, we hold the
following to be true:\bigskip
\begin{equation}
\tanh \frac{\left| J_{2}\right| }{4k_{B}T_{c}}\lessapprox
0.88,T_{c}\gtrapprox \frac{\left| J_{2}\right| }{5.5k_{B}}.
\end{equation}

\section{Summary}

\bigskip We find an exact solution to the 1D problem, in which the atom
chain ``melts'' at $0\mathrm{K}$, which is consistent with the
Lindemann criterion in the 1D case. The behavior of the atom chain
also suggests that softer lattice has greater disorder, because
for fixed temperature $T$ and ``bonding energy'' $\left|
J_{2}\right| $, both the configurational entropy (FIG.~3(a)) and
the inverse correlation length (FIG.~3(b)) increase monotonically
with the decreasing of the dimensionless ``rigidity parameter''
$\left( J_{1}+\left| J_{2}\right| \right) /\left| J_{2}\right| $
-- manifesting the reasonability of Born's criterion.

\acknowledgements

This work is supported by National Natural Science Foundation of China and
973 project.

\appendix

\section{An Graph Theoretic Approach in 1D Atom Chain}

\subsection{An Exact Graph Theoretic Approach for 1D Chain Composed of
Finite Atoms}

\smallskip The main idea of our graph theoretic solution is parallel to that
of Kac, Ward \cite{Kac}, and Vdovichenko \cite{Vdovinchenko}, as
referred to in some textbooks. (e.g. See Ref. \cite{2DIsing,
Landau}) However, the Feynman rules involved in our calculation
are non-trivial and should be carefully developed. Furthermore,
the evaluation of the thermodynamic limit requires considerable
techniques. For the completeness of this Appendix, we will
elaborate on the solution process in three steps.

\smallskip (1) We use the identity
\begin{equation}
\exp \left( \tau \theta \right) =\cosh \theta \left( 1+\tau \tanh \theta
\right) ,\tau =\pm 1
\end{equation}%
and substitutions $x_{1}=\tanh (J_{1}/4k_BT)$, $x_{2}=\tanh
(J_{2}/4k_BT) $ to transform the partition function into
\begin{equation}
Q_{N}=\left( 1-x_{1}^{2}\right) ^{-N}\left( 1-x_{2}^{2}\right)
^{-N}S\left( x_{1},x_{2}\right)
\end{equation}%
where
\begin{equation}
S\left( x_{1},x_{2}\right) =\sum_{\left\{ \sigma _{k}=\pm 1\right\}
}\prod_{k=1}^{2N}\left( 1-x_{1}\sigma _{k}\sigma _{k+1}\right) \left(
1-x_{2}\sigma _{k}\sigma _{k+2}\right)  \label{S}
\end{equation}

The summand in Eqn.~(\ref{S}) is a polynomial in the variables
$x_1,x_2$ and $\sigma _k$. Each $\sigma _k$ can appear in the
polynomial in powers ranging from zero to four. After summation
over all the $\left\{ \sigma _k=\pm 1\right\} $ states, the terms
containing odd powers of $\sigma _k$ would
vanish. Therefore, a non-zero contribution comes only from terms containing $%
\sigma _k$ in powers of $0,2$ or $4$. Since $\sigma _k^0=\sigma _k^2=\sigma
_k^4=1$, each term of the polynomial which contains all the variables $%
\sigma _k$ in even powers gives a contribution to the sum which is
proportional to the number of configurations, $2^{2N}$.

(2) We notice that each term of the polynomial can be uniquely
related to a set of lines or ``bonds'' joining various pairs of
adjoining lattice points as shown in FIG.~(\ref{lattice}).
\begin{equation}
\includegraphics[width=7cm]{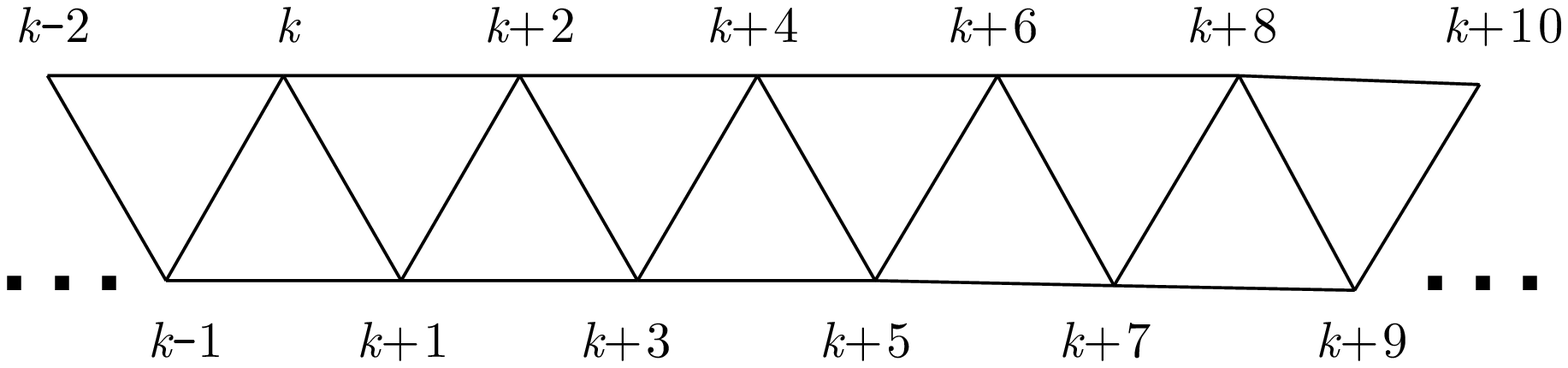}  \label{lattice}
\end{equation}

For instance, the diagrams in FIG.~(\ref{terms}) correspond to the
terms
\begin{equation*}
\begin{array}{ll}
\mathrm{(a)} & x_{1}x_{2}\sigma _{k+1}\sigma _{k}^{2}\sigma _{k+2}, \\
\mathrm{(b)} & x_{1}^{4}x_{2}^{2}\sigma _{k-2}^{2}\sigma _{k-1}^{2}\sigma
_{k}^{4}\sigma _{k+1}^{2}\sigma _{k+2}^{2}, \\
\mathrm{(c)} & -x_{1}^{4}x_{2}^{5}\sigma _{k-2}^{2}\sigma _{k-1}^{2}\sigma
_{k+1}^{2}\sigma _{k+2}^{2}\sigma _{k}^{2}\sigma _{k+4}^{2}\sigma
_{k+3}^{2}\sigma _{k+5}^{2}\sigma _{k+6}^{2}.%
\end{array}%
\end{equation*}%
\begin{equation*}
\includegraphics[width=6cm]{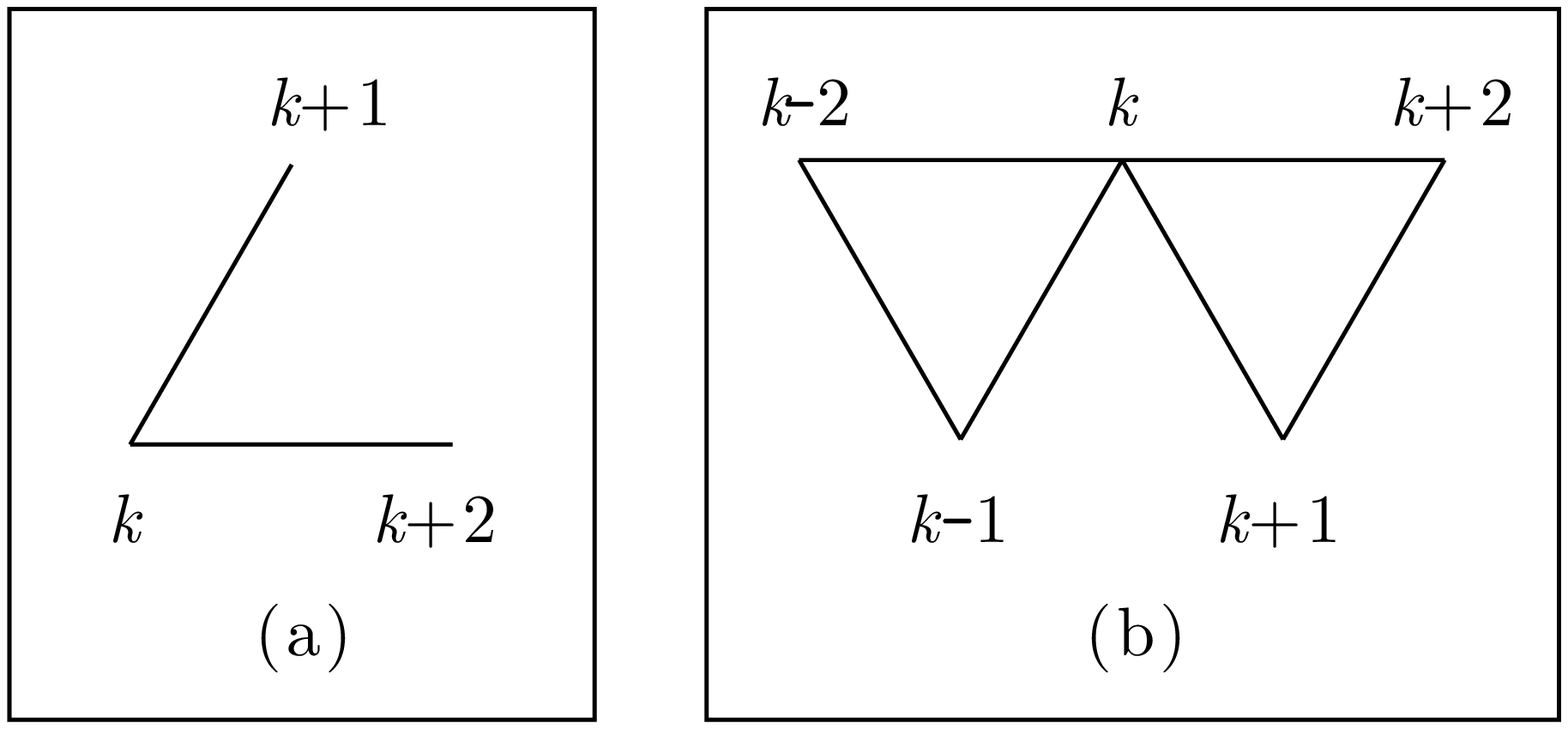}
\end{equation*}

\begin{equation}
\includegraphics[width=5cm]{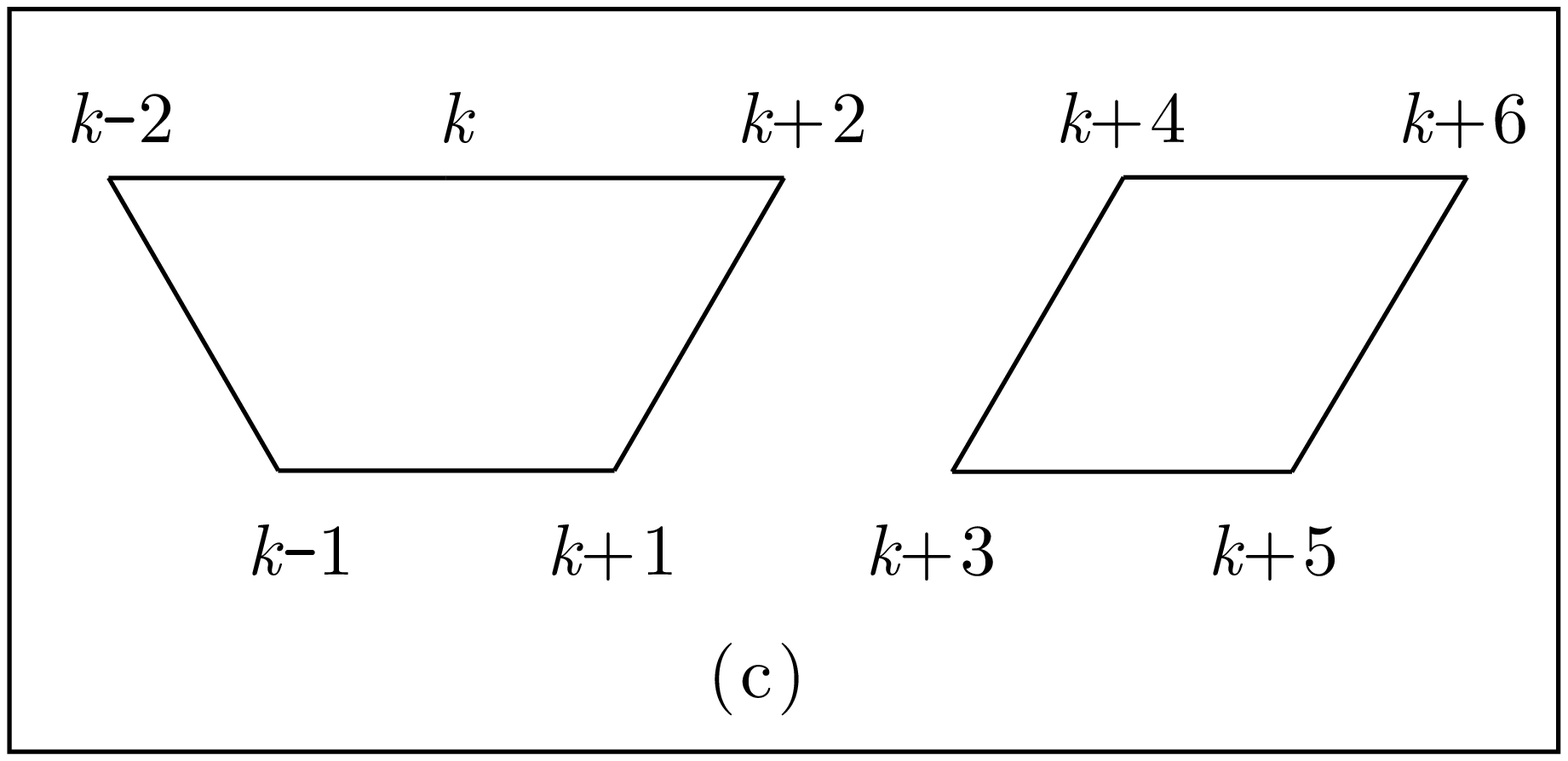}  \label{terms}
\end{equation}%
Each level line (alias ``edge'') in the diagram is assigned a
factor $\left( -x_{2}\right) $ while each oblique line a factor
$\left( -x_{1}\right) $. Each end of each line is associated with
a factor $\sigma _{k}$. The fact that a non-zero contribution to
the partition function comes only from terms in the polynomial
which contain all the $\sigma _{k}$ in even powers signifies
geometrically that either $2$ or $4$ bonds must end at each point
in the diagram.\cite{evenx} Hence the summation is taken only over
closed diagrams, which may be self-intersecting (as at the point
numbered $k$ in FIG.~(\ref{terms}) (b)).

\smallskip In the light of this, the sum $S\left( x_{1},x_{2}\right) $ may
be expressed in the form
\begin{equation}
S\left( x_{1},x_{2}\right) =2^{2N}\sum_{r}\left( -x_{2}\right)
^{r}g_{r}\left( \frac{x_{1}}{x_{2}}\right)  \label{sum}
\end{equation}%
where $g_{r}$ is a polynomial with argument $x_{1}/x_{2}$. This
polynomial represents the contributions from closed diagrams
formed from a number $r$ of bonds, with the coefficient of
$(x_{1}/x_{2})^{q}$ being equal to the number of distinct $r$-bond
long closed diagrams with an (even) number $q$ of oblique lines,
and each multiple diagram (e.g. FIG.~(\ref{terms}) (c)) being
counted once.

(3) We convert the summation over closed diagrams into one over all possible
loops, then we calculate it by a ``random walk'' method and arrive at the
following exact formula for $0<-x_2,x_1<1$:

\begin{widetext}
\textit{Theorem:}
\begin{eqnarray}
&&\frac{S\left( x_1,x_2\right) }{2^{2N}}  \notag \\
&=&\frac 1{2^{2N}}\sum_{n=0}^{2^{2N}-1}\left\{
\prod_{k=0}^{2N-1}\left[
1-x_2\left( -1\right) ^{\left( \left\lfloor \frac n{2^{k~\mathrm{mod}~%
2N}}\right\rfloor +\left\lfloor \frac n{2^{\left( k+2\right) ~\mathrm{mod}~%
2N}}\right\rfloor \right) }\right] \left. \left[ 1-x_1\left(
-1\right) ^{\left( \left\lfloor \frac
n{2^{k~\mathrm{mod}~2N}}\right\rfloor +\left\lfloor \frac
n{2^{\left( k+1\right)~\mathrm{mod}~2N}}\right\rfloor \right)
}\right]
\right\} \right.  \notag \\
&=&\left\{ \prod_{p=1}^{2N}\left[ \left( \left( 1+x_2^2\right)
\left( 1-x_1^2\right) -2x_2\left( 1-x_1^2\right) f\left( p\right)
\right) ^2+2x_1^2\left( 1-x_2\right) ^4\left( 1+f\left( p\right)
\right) \right]
\right\} ^{1/4}  \notag \\
&&  \label{product}
\end{eqnarray}
\end{widetext}
where
\begin{equation*}
f\left( p\right) =\cos \frac{\left( 2p+1-(N~\mathrm{mod}~2)\right)
\pi}N
\end{equation*}

\begin{equation*}
\left\lfloor x\right\rfloor =\text{the least integer that is
greater than or equal to }x\text{.}
\end{equation*}

\textit{Remark:} $\smallskip $Every factor in the product above is
non-negative in that it reaches the minimum value of $\left(
1+x_2\right) ^4\left( 1-x_1^2\right) ^2>0$ at $f\left( p\right)
=-1.$

We will regard each diagram as consisting of one or more closed
loops. For non-self-intersecting diagrams this is obvious; for
example, the diagram in FIG.~(\ref{terms}) (c) consists of two
loops. For self-intersecting diagrams, however, the resolution
into loops is not unique: a given diagram may consist of different
numbers of loops for different ways of construction. This is
illustrated by FIG.~(\ref{self-intersection}), which shows three
ways of representing the diagram in FIG.~(\ref{terms}) (b) as one
or two non-self-intersecting loops or as one self-intersecting
loop. (When referring to the number of intersections, we should be
heedful that every turning point should be ``polished'' before
being taken into computation.) Any intersection may similarly be
traversed in three ways on more complicated diagrams.
\begin{equation}
\includegraphics[width=7cm]{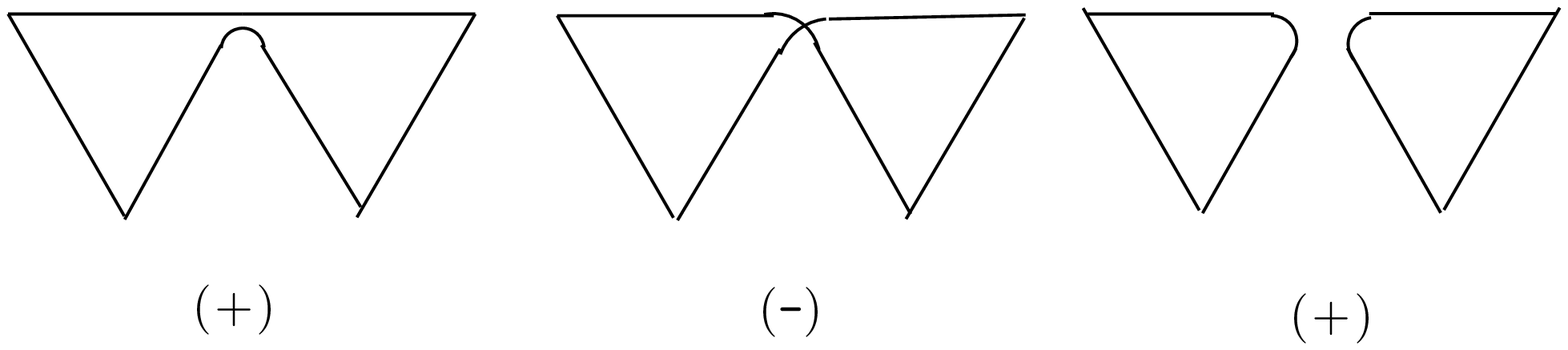}  \label{self-intersection}
\end{equation}

It is evident that the sum (\ref{sum}) can be extended to all possible sets
of loops if, in computing the contribution of diagrams to $g_{r}(x_{1}/x_{2})
$, each diagram is taken with the sign $\left( -1\right) ^{n}$, where $n$ is
the total number of self-intersections in the loops of a given set, since
when this is done all the extra terms in the sum necessarily cancel. For
example, the three diagrams in FIG.~(\ref{self-intersection}) have signs $%
+,-,+$ respectively, so that two of them cancel, leaving a single
contribution to the sum, as they should. The new sum will also
include diagrams with ``repeated bonds'', of which the simplest
example is shown in the far left of FIG.~(\ref{repeated_bonds}):
\begin{equation}
\includegraphics[width=7cm]{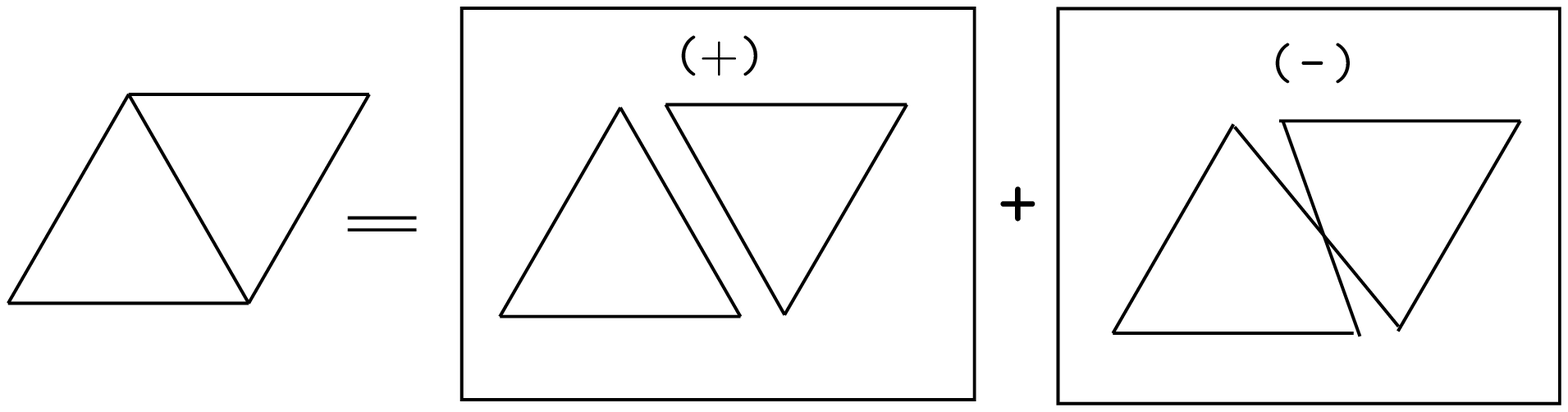}  \label{repeated_bonds}
\end{equation}%
These diagrams are not permissible, since some points have an odd
number of bonds meeting at them, namely three, but in fact they
cancel from the sum, as they should: when the loops corresponding
to such a diagram are constructed each bond in common can be
traversed in two ways, without intersection (as in the middle of
FIG.~(\ref{repeated_bonds})) and with self-intersections (far
right of FIG.~(\ref{repeated_bonds})); the resulting set of loops
appear in the sum with opposite signs, and so cancel. We can also
avoid the need to take into account explicitly the number of
intersections by using the following Feynman rules:
\begin{equation}
\includegraphics[width=7cm]{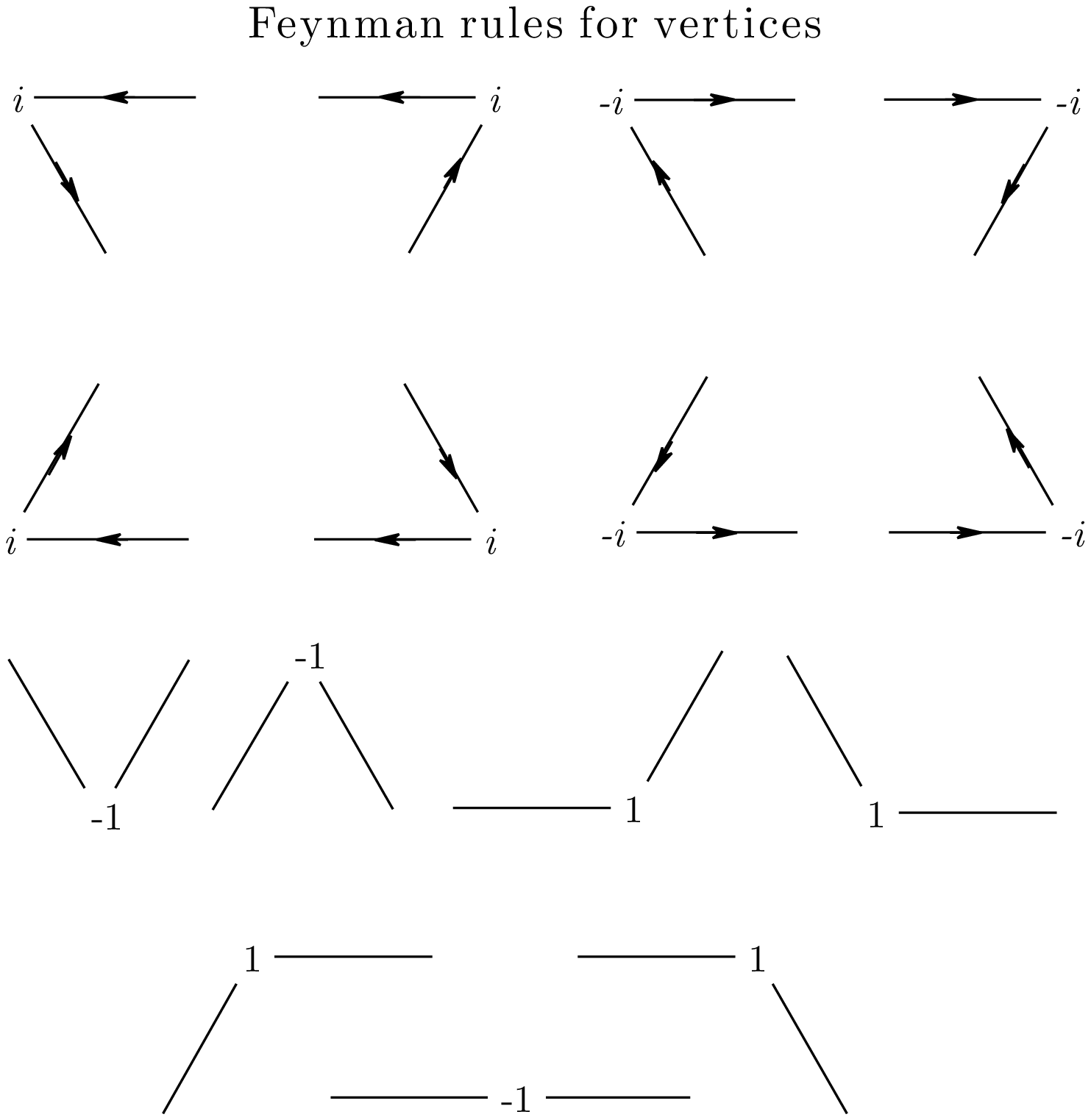}
\end{equation}%
\begin{equation}
\includegraphics[width=6cm]{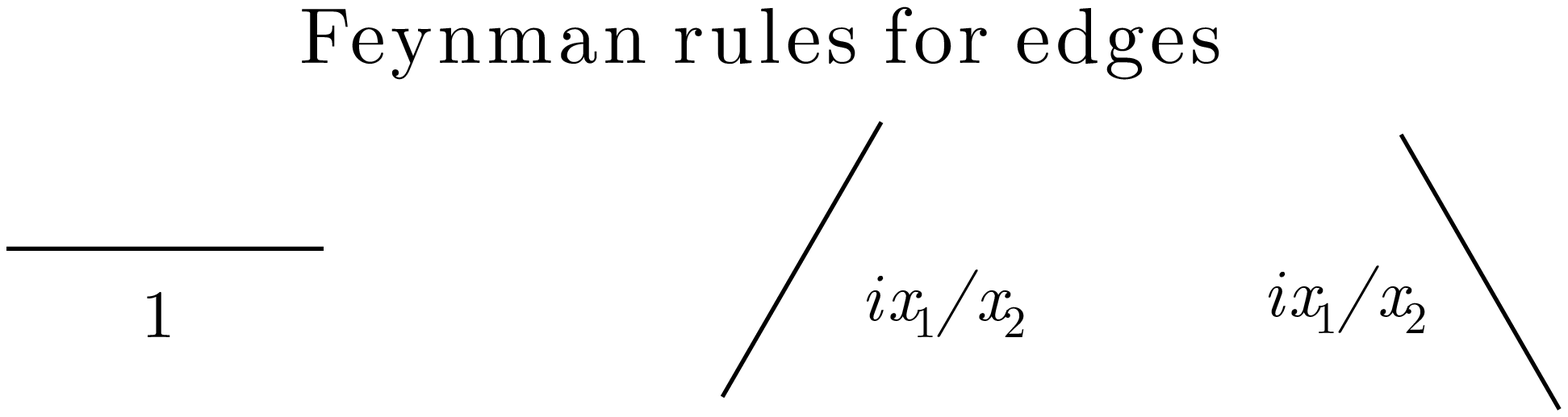}\text{\quad.}
\end{equation}

It is a direct corollary from the rules above that every
four-vertex contributes a factor $1$ in general. It can be
verified that when these rules are applied to diagrams in
FIG.~(\ref{self-intersection}), the product
of all contributions from the edges and vertices results in the sign as $%
\left( -1\right) ^{s+n}$, where $n$ is the sum of number of
self-intersections from a set of $s$ loops.\cite{rp} Here is another more
sophisticated example where contributions from the six oblique edges are not
explicitly marked:
\begin{equation}
\includegraphics[width=8cm]{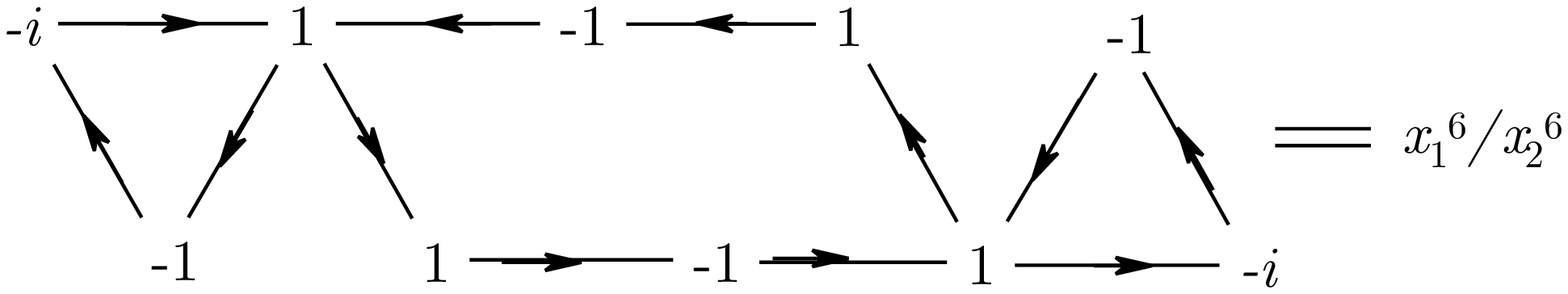}\text{.}
\end{equation}
It could be verified that the contribution of a directed graph does not
change if all the arrows are reversed. Another thing that should be
considered is that loops that ``wrap up the circle'' should bear a correct
sign. When $N$ is odd, this is automatically sufficed when the Feynman rules
above are applied. For instance, the loop that connects all the consecutive
sites together (and connects $``2N"$ to $``1"$) in a circle has the sign $%
``-"$. However, when $N$ is even, we cannot achieve a correct counting
unless we impose the following additional Feynman rules to the directed
edges:
\begin{equation}
\includegraphics[width=8cm]{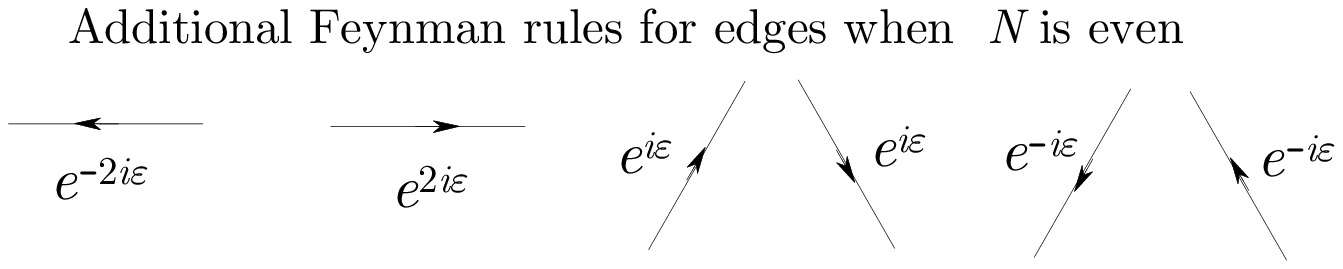}
\end{equation}
where $\varepsilon =\pi /2N$.

We use these \textit{ad hoc} Feynman rules instead of the ``geometric result
about the total angle of rotation of the tangent vector'' (a variant of
Umlaufsatz in differential geometry) as did in Ref. (\cite{Landau}), because
the latter method would result in different recurrence relations (see the
argument below) for odd number sites and even number sites, which would
plague the diagonalization procedure.

Let $f_{r}(x_{1}/x_{2})$ denote the sum over single loops of length $r$
(i.e. consisting of $r$ bonds), each loop carrying a factor $\pm 1$ as
determined by the Feynman rules, and the power of $x_{1}/x_{2}$ denoting the
number of oblique edges in that loop. Then the sum over all pairs of loops
with total number of bonds $r$ is
\begin{equation}
\frac{1}{2!}\sum_{r_{1}+r_{2}=r}f_{r_{1}}\left( \frac{x_{1}}{x_{2}}\right)
f_{r_{2}}\left( \frac{x_{1}}{x_{2}}\right) ;
\end{equation}%
the factor $1/2!$ takes into account the fact that the same pair of loops is
obtained when the suffixes $r_{1}$ and $r_{2}$ are interchanged, and
similarly for loops of three or more loops. Thus the sum becomes
\begin{eqnarray}
&&\frac{S\left( x_{1},x_{2}\right) }{2^{2N}}  \notag \\
&=&\sum_{s=0}^{\infty }\left( -1\right) ^{s}\frac{1}{s!}\sum_{r_{1},r_{2},%
\cdots =1}^{\infty }\left( -x_{2}\right) ^{r_{1}+\cdots +r_{s}}\times
\notag \\
&&f_{r_{1}}\left( \frac{x_{1}}{x_{2}}\right) \cdots f_{r_{s}}\left( \frac{%
x_{1}}{x_{2}}\right) .
\end{eqnarray}

Since $S\left( x_{1},x_{2}\right) $ includes sets of loops with every total
length $r_{1}+r_{2}+\cdots $, the numbers $r_{1},r_{2},\cdots $ in the inner
sum take independently all values from $1$ to $\infty $. Hence
\begin{eqnarray}
&&\sum_{r_{1},r_{2},\cdots =1}^{\infty }\left( -x_{2}\right) ^{r_{1}+\cdots
+r_{s}}f_{r_{1}}\left( \frac{x_{1}}{x_{2}}\right) \cdots f_{r_{s}}\left( \frac{%
x_{1}}{x_{2}}\right)  \notag \\
&=&\left( \sum_{r=1}^{\infty }\left( -x_{2}\right) ^{r}f_{r}\left( \frac{%
x_{1}}{x_{2}}\right) \right) ^{s}
\end{eqnarray}%
and $S\left( x_{1},x_{2}\right) $ becomes
\begin{equation}
\frac{S\left( x_{1},x_{2}\right) }{2^{2N}}=\exp \left( -\sum_{r=1}^{\infty
}\left( -x_{2}\right) ^{r}f_{r}\left( \frac{x_{1}}{x_{2}}\right) \right)
\label{exp}
\end{equation}

It is now convenient to assign to each lattice point (marked by an asterisk
in the figure below) the four possible directions from it and to number them
by a quantity $\nu =1,2,3,4$, say, as follows:
\begin{equation}
\includegraphics[width=6cm]{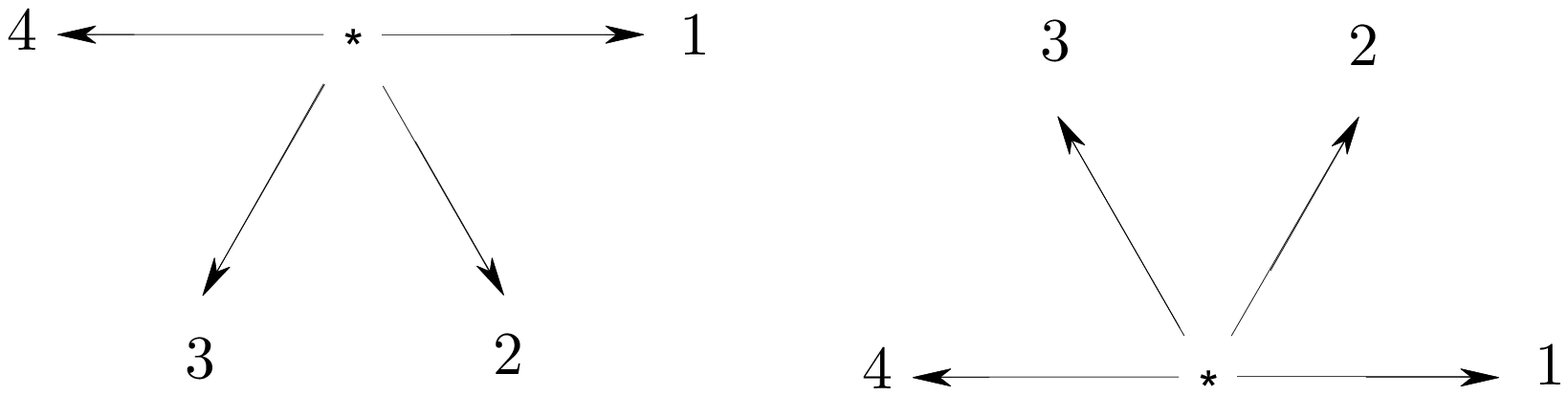}
\end{equation}

\smallskip We define an auxiliary quantity $W_{r}\left( k,\nu \right) $ the
sum over all possible paths of length $r$ from some given point $k_{0},\nu
_{0}$ to a point $k,\nu $ (each bond and vertex having as usual the factor
contributed by Feynman rules); the final step to the point $k,\nu $ must not
be from a site marked by the left-right mirror of $\nu $. (e.g. $%
1\leftrightarrow 4,2\leftrightarrow 3$) With this definition, $W_{r}\left(
k_{0},\nu _{0}\right) $ is the sum over all loops leaving the point $k_{0}$
in the direction $\nu _{0}$ and returning to that point. It is evident that
\begin{equation}
f_{r}\left( \frac{x_{1}}{x_{2}}\right) =\frac{1}{2r}\sum_{k_{0},\nu
_{0}}W_{r}\left( k_{0},\nu _{0}\right) :
\end{equation}%
both sides contain the sum over all single loops, but $\sum W_{r}$ contains
each loop $2r$ times, since it can be traversed in two opposite direction
and can be assigned to each of $r$ starting points on it.

From the definition of $W_r\left( k,\nu \right) $ we have the recurrence
relations (sometimes referred to as Chapman-Kolmogorov equations) when $N$
is an even number:

\begin{widetext}
\begin{equation} \left.
\begin{array}{l}
W_{r+1}\left( k,1\right) =-e^{2i\varepsilon }W_{r}\left( k-2,1\right) +i%
\frac{x_{1}}{x_{2}}e^{i\varepsilon }W_{r}\left( k-1,2\right) +\frac{x_{1}}{%
x_{2}}e^{-i\varepsilon }W_{r}\left( k+1,3\right) +0 \\
W_{r+1}\left( k,2\right) =e^{2i\varepsilon }W_{r}\left( k-2,1\right) -i\frac{%
x_{1}}{x_{2}}e^{i\varepsilon }W_{r}\left( k-1,2\right)
+0+ie^{-2i\varepsilon
}W_{r}\left( k+2,4\right)  \\
W_{r+1}\left( k,3\right) =-ie^{2i\varepsilon }W_{r}\left( k-2,1\right) +0-i%
\frac{x_{1}}{x_{2}}e^{-i\varepsilon }W_{r}\left( k+1,3\right)
+e^{-2i\varepsilon }W_{r}\left( k+2,4\right)  \\
W_{r+1}\left( k,4\right) =0-\frac{x_{1}}{x_{2}}e^{i\varepsilon
}W_{r}\left( k-1,2\right) +i\frac{x_{1}}{x_{2}}e^{-i\varepsilon
}W_{r}\left( k+1,3\right) -e^{-2i\varepsilon }W_{r}\left(
k+2,4\right)
\end{array}%
\right\}   \label{Chapman-Kolmogorov}
\end{equation}
\end{widetext}

The method of constructing these relations is evident: for example, the
point $k,1$ can be reached by taking the last $\left( r+1\right) $th step
from the directions marked by $1,2,3$, but not from the right (direction $4$%
); the coefficients of $W_{r}$ arise from the Feynman-rule contributions of
the last bond and vertex encountered. We can verify that these coefficients
are the same for odd number $k$ and even number $k$. Let $\Lambda $ denote
the matrix of the coefficients in Eqs. (\ref{Chapman-Kolmogorov}) (with all $%
k$), written in the form
\begin{equation*}
W_{r+1}\left( k,\nu \right) =\sum_{k^{\prime },\nu ^{\prime }}\Lambda \left(
k\nu |k^{\prime }\nu ^{\prime }\right) W_{r}\left( k^{\prime },\nu ^{\prime
}\right) .
\end{equation*}%
The method of constructing these equations enables us to associate with this
matrix an intuitive picture of a point moving step by step through the
lattice with a ``transitional probability'' per step from one point to
another which is equal to the corresponding element of the matrix $\Lambda $%
. The point traverses only one bond per step. It is evident that the
``probability'' of traversing a length $r$ will be given by the matrix $%
\Lambda ^{r}$. In particular the diagonal elements of this matrix give the
``probability'' that the point will return to its original position after
traversing a loop of length $r$, i.e. they are equal to $W_{r}\left(
k_{0},\nu _{0}\right) $. Hence
\begin{eqnarray*}
\mathrm{Tr}\Lambda ^{r} &=&\sum_{k_{0},\nu _{0}}W_{r}\left( k_{0},\nu
_{0}\right) , \\
f_{r}\left( \frac{x_{1}}{x_{2}}\right) &=&\frac{1}{2r}\mathrm{Tr}\Lambda
^{r}=\frac{1}{2r}\sum_{i}\lambda _{i}^{r},
\end{eqnarray*}%
where the $\lambda _{i}$ are the eigenvalues of the matrix.
Substituting this expression in Eqn.~(\ref{exp}) and interchanging
the order of summation over $i$ and $r$, we obtain
\begin{eqnarray}
\frac{S\left( x_{1},x_{2}\right) }{2^{2N}} &=&\exp \left\{ -\frac{1}{2}%
\sum_{i}\sum_{r=1}^{\infty }\frac{1}{r}\left( -x_{2}\right) ^{r}\lambda
_{i}^{r}\right\}  \notag \\
&=&\exp \left\{ \frac{1}{2}\sum_{i}\log \left( 1+x_{2}\lambda _{i}\right)
\right\}  \notag \\
&=&\sqrt{\prod_{i}\left( 1+x_{2}\lambda _{i}\right) }
\end{eqnarray}

\smallskip The matrix $\Lambda $ is easily diagonalized with respect to the
suffix $k$ by using a invertible transformation:

\begin{equation}
W_{r}\left( p,\nu \right) =\sum_{p=1}^{2N}e^{2ipk\varepsilon }W_{r}\left(
k,\nu \right)
\end{equation}%
where $\varepsilon =\pi /2N$. Taking ``Fourier components'' on both sides of
Eqs. (\ref{Chapman-Kolmogorov}), we find that each equation contains only $%
W_{r}(p,\nu )$ with the same $p$, so that the matrix $\Lambda $ is diagonal
with respect to $p$. For a given $p$, its elements are

\begin{widetext}
\begin{equation}
\Lambda \left( p\nu |p\nu ^{\prime }\right) =\left[
\begin{array}{cccc}
-e^{\left( 2+4p\right) i\varepsilon } &
i\frac{x_{1}}{x_{2}}e^{\left( 1+2p\right) i\varepsilon } &
+\frac{x_{1}}{x_{2}}e^{-\left( 1+2p\right)
i\varepsilon } & 0 \\
e^{\left( 2+4p\right) i\varepsilon } &
-i\frac{x_{1}}{x_{2}}e^{\left(
1+2p\right) i\varepsilon } & 0 & ie^{-\left( 2+4p\right) i\varepsilon } \\
-ie^{\left( 2+4p\right) i\varepsilon } & 0 &
-i\frac{x_{1}}{x_{2}}e^{-\left(
1+2p\right) i\varepsilon } & e^{-\left( 2+4p\right) i\varepsilon } \\
0 & -\frac{x_{1}}{x_{2}}e^{\left( 1+2p\right) i\varepsilon } & i\frac{x_{1}}{%
x_{2}}e^{-\left( 1+2p\right) i\varepsilon } & -e^{-\left(
2+4p\right)
i\varepsilon }%
\end{array}%
\right]
\end{equation}

For a given $p$, a simple calculation shows that
\begin{eqnarray}
\prod_{\nu =1}^4\left( 1+x_2\lambda _{_\nu }\right) &=&\det \left(
\delta
_{\nu \nu ^{\prime }}+x_2\Lambda _{\nu \nu ^{\prime }}\right)  \notag \\
&=&\left( 1+x_2^2\right) \left( 1-x_1^2\right) -2x_2\left(
1-x_1^2\right) \cos \frac{\left( 2p+1\right) \pi }N -2ix_1\left(
1-x_2\right) ^2\cos \frac{\left( 2p+1\right) \pi }{2N}
\end{eqnarray}
$\allowbreak $where $\lambda _{_\nu }$ is the eigenvalue of block
$\Lambda \left( p\nu |p\nu ^{\prime }\right) $. Hence,
\begin{equation*}
\Lambda \left( p\nu |p^{\prime }\nu ^{\prime }\right)
=\bigoplus_{p=1}^{2N}\Lambda \left( p\nu |p\nu ^{\prime }\right)
\end{equation*}
infers that
\begin{eqnarray}
\frac{S\left( x_1,x_2\right) }{2^{2N}} &=&\left\{
\prod_{p=1}^{2N}\left[ \left( 1+x_2^2\right) \left( 1-x_1^2\right)
-2x_2\left( 1-x_1^2\right) \cos \frac{\left( 2p+1\right) \pi }N
-2ix_1\left( 1-x_2\right) ^2\cos \frac{\left( 2p+1\right) \pi
}{2N}\right] \right\} ^{1/2}.  \label{square_root}
\end{eqnarray}
\end{widetext}
For the odd $N$ case, simply replace $2p+1$ with $2p$ and this
would yield the correct result. The product in
Eqn.~(\ref{product}) is actually obtained
by noticing that $f_r(-x_1/x_2)=f_r(+x_1/x_2)$, and by multiplying the Eqn.~(%
\ref{square_root}) by its complex conjugate.

\subsection{The Partition Function in the Thermodynamic Limit}

In the thermodynamic limit, when $N$ tends to infinity, the summation
involved in the logarithm of the partition function can be legitimately
replaced by a corresponding integral. It can be seen that both odd $N$ and
even $N$ lead to the same limit in the following integral, as they should:

\begin{eqnarray*}
& &\lim_{N\rightarrow \infty }\frac{1}{2N}\log Q_{N}  \notag \\
&=&-\frac{1}{2}\left[ \log \left( 1-x_{2}^{2}\right) +\log \left(
1-x_{1}^{2}\right) \right] +\log 2 +\frac{1}{8\pi}I(x_1,x_2) \notag \\
\end{eqnarray*}
where
\begin{widetext}
\begin{eqnarray}
 I(x_1,x_2)=\int_{0}^{2\pi
}\log \left\{ \left( \left( 1+x_{2}^{2}\right) \left(
1-x_{1}^{2}\right) -2x_{2}\left( 1-x_{1}^{2}\right) \cos 2\phi
\right) ^{2} +2x_{1}^{2}\left( 1-x_{2}\right) ^{4}\left( 1+\cos
2\phi \right) \right\} \mathrm{d}\phi
\end{eqnarray}%
The integral involved in the partition could be evaluated with
some complex analysis techniques as follows:

\textit{Lemma:} Assume that $a>0,\log 1=0$ and that $t,s$ are real
numbers, we have
\begin{eqnarray*}
I_1\left( a,t,s\right) &=&\int_0^{2\pi }\log \left[ a\left(
1+t^2+s^2+2t\left( 1+s\right) \cos m\phi +2s\cos 2m\phi \right)
\right] \mathrm{d}\phi
\\
&=&2\pi \log a
\end{eqnarray*}
where
\begin{equation}
m=1,2,\cdots ,\min_{\left| z\right| \leqslant 1}\left|
1+tz+sz^2\right| >0.
\end{equation}
\textit{Proof:}
\begin{eqnarray}
I_1\left( a,t,s\right) -2\pi \log a &=&2\mathrm{Re}\oint_{\left|
z\right|
=1}\log \left( 1+tz^m+sz^{2m}\right) \frac{\mathrm{d}z}{iz}  \notag \\
&=&2\mathrm{Re}\left( 2\pi i\mathrm{res}\left( \frac{\log \left(
1+tz^m+sz^{2m}\right) }{iz},0\right) \right) =0
\end{eqnarray}
$\blacksquare$

\textit{Corollary:}
\begin{equation}
\int_0^{2\pi }\log \left( a\left( 1+t^2+2t\cos m\phi \right)
\right) \mathrm{d}\phi =2\pi \log a,a>0,-1<t<1,m=1,2,\cdots
\end{equation}
$\blacksquare$

With these two lemmas, we are able to set out to evaluate the
integral:
\begin{eqnarray}
I_{2}\left( x_{1},x_{2}\right) &=&\int_{0}^{2\pi }\log \left\{
\left( \left( 1+x_{2}^{2}\right) \left( 1-x_{1}^{2}\right)
-2x_{2}\left( 1-x_{1}^{2}\right) \cos 2\phi \right) ^{2}
+2x_{1}^{2}\left( 1-x_{2}\right) ^{4}\left( 1+\cos 2\phi \right)
\right\}
\mathrm{d}\phi  \notag \\
&=&\int_{0}^{2\pi }\log \{ \left( 1+x_{2}^{2}\right) ^{2}\left(
1-x_{1}^{2}\right) ^{2}+2x_{1}^{2}\left( 1-x_{2}\right)
^{4}+2x_{2}^{2}\left( 1-x_{1}^{2}\right) ^{2} \notag \\ &&+\left(
-4x_{2}\left( 1+x_{2}^{2}\right) \left( 1-x_{1}^{2}\right)
^{2}+2x_{1}^{2}\left( 1-x_{2}\right) ^{4}\right) \cos 2\phi
+2x_{2}^{2}\left( 1-x_{1}^{2}\right) ^{2}\cos 4\phi \}
\mathrm{d}\phi
\end{eqnarray}
and reach the conclusion that

\textit{Theorem:}
\begin{eqnarray*}
I_2\left( x_1,x_2\right) &=&2\pi \log \left\{ \frac 12+\allowbreak
\frac 12x_2^4+2x_2^2x_1^2+x_2^2x_1^4-2x_1^2x_2-2x_1^2x_2^3
\begin{array}{l}
\! \\
{}{}\!%
\end{array}
\right. \\
&&\left.
\begin{array}{l}
\! \\
{}{}\!%
\end{array}
+\allowbreak \frac 12\left( 1-x_2\right) \left(
x_2^2-2x_1^2x_2+1\right) \sqrt{\left( 1+x_2\right)
^2-4x_1^2x_2}\right\}
\end{eqnarray*}
\begin{equation}
0\leqslant -x_2<1,0\leqslant x_1<1  \label{I2}
\end{equation}
\textit{Proof:} With
\begin{eqnarray}
A &=&\left( 1+x_2^2\right) ^2\left( 1-x_1^2\right) ^2+2x_1^2\left(
1-x_2\right) ^4+2x_2^2\left( 1-x_1^2\right) ^2  \notag \\
B &=&-4x_2\left( 1+x_2^2\right) \left( 1-x_1^2\right)
^2+2x_1^2\left(
1-x_2\right) ^4  \notag \\
C &=&2x_2^2\left( 1-x_1^2\right) ^2
\end{eqnarray}
and

\end{widetext}
\begin{eqnarray}
a\left( 1+t^2+s^2\right) &=&A>0 \\
2at\left( 1+s\right) &=&B>0 \\
2as &=&C>0,
\end{eqnarray}
we obtain
\begin{eqnarray}
a\left( 1+s+t\right) ^2 &=&A+B+C \\
a\left( 1+s-t\right) ^2 &=&A-B+C \\
\frac s{\left( 1+s\right) t} &=&\frac CB
\end{eqnarray}
and further that
\begin{equation}
\frac{\left( 1+s\right) ^2}s=\frac BC\frac{\sqrt{A+B+C}+\sqrt{A-B+C}}{\sqrt{%
A+B+C}-\sqrt{A-B+C}}
\end{equation}
\begin{widetext}

\begin{eqnarray}
a &=&\frac C{2s}  \notag \\
&=&-\frac 1{4\left( \sqrt{A-B+C}-\sqrt{A+B+C}\right) }\times  \notag \\
&&\left[ \sqrt{A-B+C}\left( B+2C\right) +\sqrt{A+B+C}\left( B-2C\right) +%
\sqrt{2}\sqrt{B^2\left( A-3C+\sqrt{A-B+C}\sqrt{A+B+C}\right)
}\right]  \notag
\\
&=&\frac 12+\allowbreak \frac
12x_2^4+2x_2^2x_1^2+x_2^2x_1^4-2x_2x_1^2-2x_2^3x_1^2+\allowbreak
\frac 12\left( 1-x_2\right) \left( x_2^2-2x_2x_1^2+1\right)
\sqrt{\left(
1+x_2\right) ^2+4x_1^2-x_2}  \notag \\
&>&0  \notag \\
&&
\end{eqnarray}
\end{widetext}

From this, it is evident that $s\left( x_1,x_2\right) =C/2a\left(
x_1,x_2\right) $ is continuous with respect to $\left( x_1,x_2\right) $ and $%
s\left( x_1,x_2\right) >0$ for $0<-x_2<1.$

We are able to affirm that $I_{2}\left( x_{1},x_{2}\right) =2\pi \log
a\left( x_{1},x_{2}\right) $ $\left( 0<-x_{2}<1,0<x_{1}<1\right) $ if we can
further that $\min_{\left| z\right| \leqslant 1}\left| 1+t\left(
x_{1},x_{2}\right) z+s\left( x_{1},x_{2}\right) z^{2}\right| >0$ is true for
all $\left( x_{1},x_{2}\right) $ such that $0<-x_{2}<1,0<x_{1}<1.$ To verify
this, we notice that for $0<-x_{2}<1$, $a\left( 0,x_{2}\right) =1,t\left(
0,x_{2}\right) =-2x_{2},s\left( 0,x_{2}\right) =2x_{2}^{2},$ so it is easy
to find that $\min_{\left| z\right| \leqslant 1}$ $\left| 1+t\left(
0,x_{2}\right) z+s\left( 0,x_{2}\right) z^{2}\right| =\min_{\left| z\right|
\leqslant 1}$ $\left| 1-x_{2}z\right| ^{2}>0.$ In other words, the roots of
the equation $1+tz+sz^{2}$ lie outside the unit disk $\left( \left| z\right|
\leqslant 1\right) $ when $-x_{2}=1/2,x_{1}=0$. For an arbitrary point $%
\left( x_{1}^{\ast },x_{2}^{\ast }\right) $ that satisfies $0<-x_{2}^{\ast
}<1,0<x_{1}^{\ast }<1$, we may connect it to the point $\left( 0,1/2\right) $
with a line segment. It is obvious that the two roots of $1+t\left(
x_{1},x_{2}\right) z+s\left( x_{1},x_{2}\right) z^{2}$ vary continuously as $%
\left( x_{1},x_{2}\right) $ moves along this line segment. The orbits of the
two roots must be two continuous paths in the complex plane in the process
mentioned above, and each of them begins with a point outside the unit disk
and cannot end up with a point inside the unit disk unless it hits the unit
circle $\left( \left| z\right| =1\right) $ some time. However, neither root
can hit the unit circle because it is true that $\min_{0\leqslant \theta
<2\pi }$ $\left| 1+t\left( x_{1},x_{2}\right) e^{i\theta }+s\left(
x_{1},x_{2}\right) e^{2i\theta }\right| =\left( 1+x_{2}\right) ^{2}\left(
1-x_{1}^{2}\right) /\sqrt{a\left( x_{1},x_{2}\right) }>0$. Therefore, both
roots of the equation $1+t\left( x_{1}^{\ast },x_{2}^{\ast }\right)
z+s\left( x_{1}^{\ast },x_{2}^{\ast }\right) z^{2}$ should lie outside the
unit disk. This completes the proof of $I_{2}\left( x_{1},x_{2}\right) =2\pi
\log a\left( x_{1},x_{2}\right) $ $\left( 0<-x_{2}<1,0<x_{1}<1\right) $, and
the generalization to $x_{2}=0$ or $x_{1}=0$ case is a trivial calculation.

$\blacksquare$

It can be verified by simple algebra that Eqn.~(\ref{I2}) gives
the free energy expression that is exactly equivalent to
Eqn.~(\ref{F}). This circuitous approach provides an independent
method to evaluate the partition function in the 1D atom chain.

\bigskip

$\dagger $ Present Address: Department of Chemistry and Chemical Biology,
Harvard University, Cambridge, MA 02138, USA.

$\ddagger $ To whom correspondence should be addressed. Email:
xfjin@fudan.ac.cn


\begin{thebibliography}{99}
\bibitem{3DJ1} It should be mentioned here that in 3D case, near the melting
point, we could only expect $J_{1}>J_{2}$ and $J_{2}<0$, but could no longer
suppose that $J_{1}>0$. This is because (1) at finite temperature, $%
J_{1}>J_{2}$ is already sufficient to guarantee that NNN\ atom pairs are
more stable than NN\ atom pairs; (2) experimental results about the Frenkel
defect formation energy in fcc metals are $(6J_{1}-12J_{2})\approx 4\sim 5$%
eV, while the bonding energy $-J_{2}=0.5\sim 1$eV, so the
$J_{1}<0$ case will be inevitable while modelling a realistic
solid in three dimensions.
\bibitem{1DJ1} Fortunately, most exact results in this paper do not change
when $J_{1}$ is replaced by $-J_{1}$, for instance, $F_{\text{conf}}$, $S_{%
\text{conf}}$, $\left\langle \sigma _{k}\sigma _{k+2}\right\rangle $, $%
\partial c/\partial T$ $\cdots $, but $\left\langle \sigma _{k}\sigma
_{k+1}\right\rangle $ and $\left( \partial S_{\text{conf}}/\partial
J_{1}\right) _{J_{2},T}$ always bear the same sign as $-J_{1}$. For a
negative NN pair energy $J_{1}$, the correlation length (represented by $\mu
\left( J_{1},J_{2},T\right) $) equals the one (represented by $\lambda
\left( \left| J_{1}\right| ,J_{2},T\right) $) for positive NN pair energy $%
\left| J_{1}\right| $. However, the form of the correlation functions are
generally different for positive and negative NN\ pair energies, $\partial
\mu \left( J_{1},J_{2},T\right) /\partial \left| J_{1}\right| =\partial
\lambda \left( \left| J_{1}\right| ,J_{2},T\right) /\partial \left|
J_{1}\right| $.
\bibitem{evenx} Since each oblique edge that links a lattice site to an interstitialcy site changes
the parity of the site label, a closed diagram should include an
even number of oblique edges. Therefore, the power of $x_1$ should
be an even number for the diagram contribution to be non-zero.
That is why the partition function is invariant under the
transformation $J_1\mapsto -J_1$.
\bibitem{rp} This result can be indeed generalized rigorously to all
diagrams, including those with ``repeated-bonds''.
\end{thebibliography}
\end{document}